\documentstyle{article} 

\def\QOVERD#1#2#3#4{{#3 \overwithdelims#1#2 #4}}
\def\stackunder#1#2{\mathrel{\mathop{#2}\limits_{#1}}}
\def\limfunc#1{\mathop{\rm #1}}

\begin{document}

\title{Quantum chaos in the framework of complex probability processes}
\author{Alexander V. Bogdanov and Ashot S. Gevorkyan}
\date{Institute for High-Performance Computing and Data Bases \\
P/O Box 71, 194291, St-Petersburg, RUSSIA \\
bogdanov@hm.csa.ru, ashot@fn.csa.ru}

\maketitle

\begin{abstract}
The problem of random motion of 1D quantum reactive harmonic oscillator
(QRHO) is formulated in terms of a wave functional regarded as a complex
probability process $\Psi _{stc}\left( x,t;\left\{ \xi \right\} \right) $
in an extended space $\Xi =R^1\otimes R_{\left\{ \xi \right\} }$. In the
complex stochastic differential equation (SDE) for $\Psi _{stc}\left(
x,t;\left\{ \xi \right\} \right) $ the variables are separated with the help
of the Langevin-type model SDE introduced in the functional space $%
R_{\left\{ \xi \right\} }$. The complete positive Fokker-Plank measure of
the space $R_{\left\{ \xi \right\} }$ is obtained. The average wave function
of roaming QRHO is obtained by means of functional integration over the process
$\Psi _{stc}\left( x,t;\left\{ \xi \right\} \right) $ with the complete
Fokker-Plank measure in the space $R_{\left\{ \xi \right\} }$. The local and
averaged transition matrices of roaming QRHO were constructed. The
thermodynamics of nonrelativistic vacuum is investigated in detail and 
expressions for the internal energy, Helmholtzian energy and entropy are
obtained. The oscillator's ground state energy, its shift and broadening
are calculated.
\end{abstract}


\section{Introduction}

From the point of view of classical dynamics all phenomena, that are
described in the framework of conventional theory of quantum mechanics are
stochastic processes. The natural symmetry between the Schr\"{o}dinger and
Fokker-Plank equations was used both for the formulation of quantum
mechanics as a classical stochastic theory \cite{Nelson} and quantization of
classical theory by introduction of the concept of random process \cite{Parisi}.

Note, that in all representations the main object of quantum mechanics, the
wave function, is deterministic. It is worthwhile to stress that the
deterministic nature of a physical theory is a consequence of the symmetry
of basic equations with respect to time inversion \cite{Gutzwiller}.

It is known that the observability is an essential property of objects that
is realized when the contact with an external world is established.
Therefore, all objects under study are open physical systems. This
conclusion implies, in particular, that it is impossible to conform in any
reasonable way the probabilistic, irreversible nature of measurements to the
reversibility of quantum theory concepts \cite{Von Oppen}.

Moreover, the stochastic nature is the basic property of the physical world.
This assertion becomes especially conclusive if one allows for the real
existence of physical vacuum in the nature that is the cause of casual quantum
jumps in the isolated systems \cite{Birrell}. In other words, all objects that
are investigated taking into account the interaction with fluctuations of the
physical vacuum are open physical systems. Whereas the isolated systems are
studied by means of methods of conventional quantum mechanics, the behaviour
of open systems, in particular of systems that are in equilibrium with the
environment, is usually described by the laws of thermodynamics.

Since due to permanent fluctuations the open system has no any definite
quantum state, i.e., the state vector of the system randomly changes with
time, the construction of a theory alternative to the reversible quantum
mechanics becomes urgent. One can construct such a theory if, in particular,
the random processes taking place in the system are directly included in the
foundations of the theory and are not regarded as external perturbations.

In the present work the problem of stochastic quantum mechanics is
investigated using the model of random motion of 1D quantum reactive
oscillator (QRHO) . Mathematically the problem of an evolution of quantum
states is formulated in the framework of probabilistic complex process in an
extended space with stochastic space-time continuum. For roaming classical
oscillator the conditions of reducibility of a model stochastic differential
equation (SDE) to nonlinear Langevin SDE are investigated. With the help of
model SDE the variables in the SDE for a wave functional - the complex
probability process, have been separated. Based on the Langevin-type SDE
the Fokker-Plank equation is derived and the measure of functional space is
obtained. For roaming QRHO the average wave function is constructed in the
form of functional integral with positive measure. The local stochastic and
averaged transition matrix of the roaming QRHO is calculated analytically.
The probability of "vacuum-vacuum" transition is calculated numerically and
its behavior is shown versus the fluctuation parameter.
The thermodynamics of nonrelativistic vacuum is builded and the expressions
for the main thermodynamic potentials are obtained. 


\section{Formulation of the problem}

Let us consider a closed system "quantum object+thermostat" and assume
that the quantum object moves in the Euclidean space $R^1$ and the state of
the thermostat is characterized by an unlimited set of modes in the
functional space $R_{\left\{ \xi \right\} }$, where $\left\{ \xi \right\}
\equiv \xi \left( t,\left\{ W\right\} \right) $ is a complex functional of
some random process $\left\{ W\right\} \equiv W\left( t\right) $. One can
mathematically describe the wave state of the system with the help of a
functional - a complex probabilistic process $\Psi _{stc}\left( x,t;\left\{
\xi \right\} \right) $, given on the extended space $\Xi =R^1\otimes
R_{\left\{ \xi \right\} }$. The equation of evolution for the wave
functional can be written in the most general case as:

\begin{equation}
\label{eq2.1}
id_t\Psi _{stc}=\hat{H}\left( x,t;\left\{ W\right\} \right) \Psi
_{stc},\qquad -\infty <x,t<+\infty ,
\end{equation}

where $d_t\equiv \left. d\right/ dt$ is the total derivative of the random
functional (see (\ref{eq3.2})) with respect to time, and $\hat{H}\left(
x,t;\left\{ W\right\} \right) $ is respectively the stochastic operator
of the system evolution.

Below, as a specific example, we shall consider the case of roaming 1D QRHO,
the operator of evolution of which has the following form:

\begin{equation}
\label{eq2.2}
\hat H\left( x,t;\left\{ W\right\} \right) =\frac 12\left[ -\partial
_x^2+\Omega ^2\left( t;\left\{ W\right\} \right) x^2\right] ,\qquad \partial
_x^2\equiv \left. \partial ^2\right/ \partial x^2.
\end{equation}

Let the frequency be given as:

\begin{equation}
\label{eq2.3}
\Omega \left( t;\left\{ W\right\} \right) =\Omega _0\left( t\right) +\Omega
_1\left( t;\left\{ W\right\} \right) ,
\end{equation}

where the regular part $\Omega _0\left( t\right) >0$ satisfies the following
boundary conditions:

\begin{equation}
\label{eq2.4}
\mathrel{\mathop{\lim }\limits_{t\rightarrow \mp \infty }}\Omega _0\left(
t\right) =\Omega _{in\left( out\right) }>0.
\end{equation}

Note, that when $\Omega _1\left( t;\left\{ W\right\}
\right) \equiv 0$, the equations (\ref{eq2.1})-(\ref{eq2.2}) describes the
case of regularly moving QRHO with asymptotic wave function (\ref{eq2.7}),
that admits exact solution (see \cite{Bjaz}). In the expression (\ref{eq2.3})
$\Omega _1\left( t;\left\{ W\right\} \right) $ respectively denotes the
random part of frequency and satisfies the following conditions:

\begin{equation}
\label{eq2.5}
\mathrel{\mathop{\lim }\limits_{t\rightarrow -\infty }}\Omega _1\left(
t;\left\{ W\right\} \right) =0,\qquad \mathrel{\mathop{\lim
}\limits_{t\rightarrow +\infty }}\Omega _1\left( t;\left\{ W\right\} \right)
\neq 0.
\end{equation}

Next, assume that the total wave functional meets the natural boundary
conditions:

\begin{equation}
\label{eq2.6}
\mathrel{\mathop{\lim }\limits_{\left| x\right| \rightarrow \infty }}\Psi
_{stc}\left( x,t;\left\{ \xi \right\} \right) =\mathrel{\mathop{\lim
}\limits_{\left| x\right| \rightarrow \infty }}\partial _x\Psi _{stc}\left(
x,t;\left\{ \xi \right\} \right) =0.
\end{equation}

We denote as $\Psi _{stc}^{\left( +\right) }\left( n|x,t;\left\{ \xi
\right\} \right) $ the total wave functional that is evolved in the
Euclidean subspace $R_{in}^1$ from the purely $n$-th vibrational state of
the quantum harmonic oscillator:

$$
\Psi _{stc}^{\left( +\right) }\left( n|x,t;\left\{ \xi \right\} \right) %
\mathrel{\mathop{\rightarrow }\limits_{t\rightarrow -\infty }}\Psi
_{in}\left( n|x,t\right) =
$$

\begin{equation}
\label{eq2.7}
=\left[ \frac{\left( \left. \Omega _{in}\right/ \pi \right) ^{\frac12 }}{%
2^nn!}\right] ^{1/2}\exp \left\{ -i\left( n+\frac 12\right) \Omega
_{in}t-\frac 12\Omega _{in}x^2\right\} H_n\left( \sqrt{\Omega _{in}}x\right) ,
\end{equation}

$$
n=0,1,2...
$$

The objectives of the present investigation are:

a) to establish the criteria for separation of variables in SDE (\ref{eq2.1}%
)-(\ref{eq2.2}) and determine the random wave functional $\Psi
_{stc}^{\left( +\right) }\left( n|x,t;\left\{ \xi \right\} \right) $ in an
explicit form;

b) to calculate the wave function of roaming QRHO

\begin{equation}
\label{eq2.8}
\Psi _{br}^{\left( +\right) }\left( n|x,t\right) =\left\langle \Psi
_{stc}^{\left( +\right) }\left( n|x,t;\left\{ \xi \right\} \right)
\right\rangle _{\left\{ \xi \right\} },
\end{equation}

where the brackets $\left\langle \ldots \right\rangle _{\left\{ \xi \right\}
}$ stand for the functional integration including the integration over the
distribution of coordinate $\xi $ in the functional space $R_{\left\{ \xi
\right\} }$ at instant $t$;

c) to calculate the transition matrix $S_{nm}^{br}$ of the roaming QRHO.


\section{ Derivation of SDE for roaming classical oscillator}

We begin our consideration with a second-order equation

\begin{equation}
\label{eq3.1}
\ddot{\xi}+\Omega ^2\left( t;\left\{ W\right\} \right) \xi =0,
\end{equation}

where

\begin{equation}
\label{eq3.2}
\dot \xi =d_t\xi \left( t;\left\{ W\right\} \right) ,\qquad d\xi \left(
t;\left\{ W\right\} \right) =\left( \partial _t\xi +\frac 12\delta _W^2\xi
\right) dt+\left( \delta _W\xi \right) dW\left( t\right) .
\end{equation}

Remember that the second formula in (\ref{eq3.2}) denotes the Ito
differential of the random functional $\xi \left( t;\left\{ W\right\}
\right) $ (see \cite{Gardiner}), where the functional derivative is determined
in the usual way:

\begin{equation}
\label{eq3.3}
\delta _W\xi =\left\{ \frac{\delta \xi \left( t;W\left( t\right) \right) }{%
\delta W\left( t^{\prime }\right) }\right\} _{t=t^{\prime }}.
\end{equation}

Taking into account the conditions (\ref{eq2.3})-(\ref{eq2.4}) we find the
asymptotical solution of (\ref{eq3.1}) in the $(in)$ channel to be

\begin{equation}  
\label{eq3.4}
\xi \left( t;\left\{ W\right\} \right) \mathrel{\mathop{\sim
}\limits_{t\rightarrow -\infty }}\exp \left( i\Omega _{in}t\right) .
\end{equation}

One is to stress here, that the equation (\ref{eq3.1}) makes sense and
describes the behaviour of roaming classical oscillator only when it is
reducible to the Langevin-type SDE.

{\bf Theorem:} {\it There exists a set of nonsingular functionals $\xi
\left( t;\left\{ W\right\} \right) $ that reduce the equation (\ref{eq3.1})
to a nonlinear complex Langevin-type SDE, or, that is the same, to a system
of two real nonlinear SDE.}

{\it Proof:} Assume that the solution of model equation (\ref{eq3.1}) has
the following form:

\begin{equation}
\label{eq3.5}
\xi \left( t;\left\{ W\right\} \right) =\xi _0\left( t\right) \exp \left( 
\stackrel{t}{\mathrel{\mathop{\int }\limits_{-\infty }}}\phi \left(
t^{\prime };\left\{ W^{\prime }\right\} \right) dt^{\prime }\right) ,
\end{equation}

where $\xi _0\left( t\right) $ is the solution of (\ref{eq3.1}) with regular
frequency $\Omega _0(t)$ (see \cite{Bjaz}). Consider the first total derivative
of the random functional (\ref{eq3.5}) with respect to time. Taking into
account (\ref{eq3.2}) one has

$$
d_t\xi \left( t;\left\{ W\right\} \right) =\left[ \left. \xi _{0t}\left(
t\right) \right/ \xi _0\left( t\right) +\phi \left( t;\left\{ W\right\}
\right) +\right. 
$$

\begin{equation}  
\label{eq3.6}
\left. +\frac 12\stackrel{t}{\mathrel{\mathop{\int }\limits_{-\infty }}}%
\delta _{W}^2\phi \left( t^{\prime };\left\{ W\right\} \right) dt^{\prime
}+\left( d_tW\left( t\right) \right) \stackrel{t}{\mathrel{\mathop{\int
}\limits_{-\infty }}}\delta _{W^{\prime }}\phi \left( t^{\prime };\left\{
W^{\prime }\right\} \right) dt^{\prime }\right] \xi \left( t;\left\{
W\right\} \right) ,
\end{equation}

$$
\xi _{0t}\left( t\right) =d_t\xi _0\left( t\right) . 
$$

Since for all $t^{\prime }<t$

\begin{equation}
\label{eq3.7}
\left. \delta _{W^{\prime }}\phi \left( t^{\prime };\left\{ W^{\prime
}\right\} \right) \right| _{t^{\prime }<t}=0,  
\end{equation}

we can rewrite the expression (\ref{eq3.6}) as

\begin{equation}  
\label{eq3.8}
d_t\xi \left( t;\left\{ W\right\} \right) =\left[ \left. \xi _{0t}\left(
t\right) \right/ \xi _0\left( t\right) +\phi \left( t;\left\{ W\right\}
\right) \right] \xi \left( t;\left\{ W\right\} \right) =\partial _t\xi
\left( t;\left\{ W\right\} \right) .
\end{equation}

Substituting the solution (\ref{eq3.5}) to (\ref{eq3.1}) and taking into
account (\ref{eq3.8}) one can find:

\begin{equation}
\label{eq3.9}
\dot{\phi}+2\left[ \left. \xi _{0t}\left( t\right) \right/ \xi _0\left(
t\right) \right] \phi +\phi ^2+\Omega _o^2\left( t\right) +U\left( t;\left\{
W\right\} \right) =0,  
\end{equation}

where the notation was made

\begin{equation}  
\label{eq3.10}
U\left( t;\left\{ W\right\} \right) =\Omega _1^2\left( t;\left\{ W\right\}
\right) +2\Omega _1\left( t;\left\{ W\right\} \right) \Omega _0\left(
t\right) .
\end{equation}

In particular, from (\ref{eq3.7}) and equation (\ref{eq3.9}) the following
relations results

\begin{equation}
\label{eq3.11}
\delta _W\xi \left( t;\left\{ W\right\} \right) =0,\qquad \delta _W\left\{
\partial _t\xi \left( t;\left\{ W\right\} \right) \right\} \neq 0
\end{equation}

from which it is seen that the operators $\delta _W$ and $\partial _t$ do
not commute.

Making the substitution

\begin{equation}
\label{eq3.12}
\phi \left( t;\left\{ W\right\} \right) =\Phi \left( t;\left\{ W\right\}
\right) -\left. \xi _{0t}\left( t\right) \right/ \xi _0\left( t\right) ,
\end{equation}

one obtain finally from (\ref{eq3.9})

\begin{equation}  
\label{eq3.13}
\dot \Phi +\Phi ^2+\Omega _0^2\left( t\right) +F_0+F\left( t;\left\{
W\right\} \right) =0.
\end{equation}

In SDE (\ref{eq3.13}) the following notations were made

\begin{equation}
\label{eq3.14}
U\left( t;\left\{ W\right\} \right) =F_0+F\left( t;\left\{ W\right\} \right)
,\qquad F_0=\left\langle U\left( t;\left\{ W\right\} \right) \right\rangle ,  
\end{equation}

and in (\ref{eq3.14}) the averaging of expression (\ref{eq3.10}) is made
over the ensemble.

The solution (\ref{eq3.13}) of SDE, taking into account (\ref{eq3.4}),
satisfies the initial condition of the type

\begin{equation}
\label{eq3.15}
\mathrel{\mathop{\lim }\limits_{t\rightarrow -\infty }}\Phi \left( t;\left\{
W\right\} \right) =i\Omega _{in},  
\end{equation}

that, in its turn, suggests the complex solution

\begin{equation}  
\label{eq3.16}
\Phi \left( t;\left\{ W\right\} \right) =\theta \left( t;\left\{ W\right\}
\right) +i\varphi \left( t;\left\{ W\right\} \right) .
\end{equation}

After substitution of (\ref{eq3.16}) to (\ref{eq3.13}) and separation of
real and imaginary parts we have the following system of SDE:

\begin{equation}  
\label{eq3.17}
\dot \theta +\theta ^2-\varphi ^2+U_0\left( t\right) +F\left( t;\left\{
W\right\} \right) =0,\qquad U_0\left( t\right) =\Omega _0^2\left( t\right)
+F_0>0,
\end{equation}

\begin{equation}  
\label{eq3.18}
\dot \varphi +2\varphi \theta =0.
\end{equation}

The system of SDE (\ref{eq3.17})-(\ref{eq3.18}) can be rewritten in a vector
equation:

\begin{equation}
\label{eq3.19}
\stackrel{.}{\bf \Phi }+{\bf K}\left( t;\left\{ {\bf \Phi }\right\} \right) +%
{\bf F}\left( t;\left\{ W\right\} \right) =0,  
\end{equation}

where the vectors have the following projections

$$
{\bf \Phi }\left( t;\left\{ W\right\} \right) =\left\{ \theta ,\varphi
\right\} ,\qquad {\bf K}\left( t;\left\{ {\bf \Phi }\right\} \right)
=\left\{ \left[ \theta ^2-\varphi ^2+U_0\left( t\right) +F_0\right] ;2\theta
\right\} , 
$$

\begin{equation}  
\label{eq3.20}
{\bf F}\left( t;\left\{ W\right\} \right) =\left\{ F\left( t;\left\{
W\right\} \right) ;0\right\} .
\end{equation}

{\it The theorem is proved.}

For subsequent investigations of SDE (\ref{eq2.1})-(\ref{eq2.2}) it is
convenient to write SDE (\ref{eq3.1}) in the form

\begin{equation}
\label{eq3.21}
\xi \left( t;\left\{ W\right\} \right) =\sigma \left( t;\left\{ W\right\}
\right) \exp \left[ ir\left( t;\left\{ W\right\} \right) \right] ,
\end{equation}

where $\sigma \left( t;\left\{ W\right\} \right) $ and $r\left( t;\left\{
W\right\} \right) $ are yet unknown real functionals:

\begin{equation}  
\label{eq3.22}
\mathop{\rm Im}\sigma \left( t;\left\{ W\right\} \right) =\mathop{\rm Im}%
r\left( t;\left\{ W\right\} \right) =0.
\end{equation}

Taking into account (\ref{eq3.4}) one can establish the asymptotic behaviour
of these functionals:

\begin{equation}  
\label{eq3.23}
\mathrel{\mathop{\lim }\limits_{t\rightarrow -\infty }}\sigma \left(
t;\left\{ W\right\} \right) =1,\qquad r\left( t;\left\{ W\right\} \right) %
\mathrel{\mathop{\rightarrow }\limits_{t\rightarrow -\infty }}\Omega _{in}t.
\end{equation}

After substitution of (\ref{eq3.21}) to (\ref{eq3.11}) and taking into account
(\ref{eq3.16}), (\ref{eq3.18}) and (\ref{eq3.22}) we find that

\begin{equation}
\label{eq3.24}
\delta _W\sigma \left( t;\left\{ W\right\} \right) =0,\qquad \delta _W\left[
\partial _t\sigma \left( t;\left\{ W\right\} \right) \right] =0,
\end{equation}

as well as that

\begin{equation}  
\label{eq3.25}
\delta _Wr\left( t;\left\{ W\right\} \right) =0,\qquad \delta _W\left[
\partial _tr\left( t;\left\{ W\right\} \right) \right] =0.
\end{equation}

It must be noted that the conditions (\ref{eq3.24})-(\ref{eq3.25}) are
highly important for our further discussions.


\section{Solution of SDE for complex probabilistic process - the wave
functional}

Now pass to the solution of SDE (\ref{eq2.1})-(\ref{eq2.2}) using the
Langevin-type model nonlinear SDE for separation of variables.

{\bf Theorem:} {\it If the relations (\ref{eq3.24})-(\ref{eq3.25}) are fulfilled,
then the complex SDE (\ref{eq2.1})-(\ref{eq2.2}) has an exact solution in
the form of orthonormal in $L_2\left( R^1\otimes R_{\left\{ \xi \right\}
}\right) $ random complex functionals 

$$
\Psi _{stc}^{\left( +\right) }\left( n|x,t;\left\{ \xi \right\} \right)
=\left[ \frac{\left( \left. \Omega _{in}\right/ \pi \right) ^{1/2}}{%
2^nn!\sigma }\right] ^{1/2}\exp \left\{ -i\left( n+\frac 12\right) \Omega
_{in}\tau +\right. 
$$

\begin{equation}
\label{eq4.1}
\left. +i\frac{\sigma _t}{2\sigma }x^2-\frac 12r_tx^2\right\} H_n\left( 
\sqrt{\Omega _{in}}\frac x\sigma \right) ,  
\end{equation}

where $\tau \left( t;\left\{ W\right\} \right) =\left. r\left( t;\left\{
W\right\} \right) \right/ \Omega _{in}$, $\sigma _t=\partial _t\sigma \left(
t;\left\{ W\right\} \right) $ and $r_t=\partial _tr\left( t;\left\{
W\right\} \right) $. Moreover, $L_2\left( R^1\otimes R_{\left\{ \xi \right\}
}\right) $ denotes the space of square integrable functions on the real
axis $R^1 $ with values in some functional space $R_{\left\{ \xi \right\} }$.}

{\it Proof:} Changing the variables in (\ref{eq2.1})-(\ref{eq2.2})

\begin{equation}
\label{eq4.2}
x\rightarrow y=\frac x{\sigma \left( t;\left\{ W\right\} \right) },
\end{equation}

and taking into account (\ref{eq3.24})-(\ref{eq3.25}) the equation for
complex stochastic process will be as follows:

$$
id_t\tilde \Psi _{stc}\left( y,t;\left\{ \xi \right\} \right) =\left\{ i%
\frac{\sigma _t}\sigma y\delta _y-\frac 1{2\sigma ^2}\delta _y^2+\frac{%
\sigma ^2}2\Omega ^2\left( t;\left\{ W\right\} \right) y^2\right\} \tilde
\Psi _{stc}\left( y,t;\left\{ \xi \right\} \right) , 
$$

\begin{equation}  
\label{eq4.3}
\tilde \Psi _{stc}\left( y,t;\left\{ \xi \right\} \right) =\Psi _{stc}\left(
x,t;\left\{ \xi \right\} \right) .
\end{equation}

Now, let us rewrite the solution of equation (\ref{eq4.3}) as

\begin{equation}  
\label{eq4.4}
\tilde \Psi _{stc}\left( y,t;\left\{ \xi \right\} \right) =\left\{ \frac{%
\exp \left[ i2\Lambda \left( t;\left\{ W\right\} \right) y^2\right] }{\sigma
\left( t;\left\{ W\right\} \right) }\right\} ^{1/2}\chi \left( y,\tau \left(
t;\left\{ W\right\} \right) \right) .
\end{equation}

Making the substitution of (\ref{eq4.4}) into (\ref{eq4.3}) and the
transformation

\begin{equation}
\label{eq4.5}
t\rightarrow \tau =\frac{r\left( t;\left\{ W\right\} \right) }{\Omega _{in}},
\end{equation}

with regard to (\ref{eq3.24})-(\ref{eq3.25}) one finds

$$
i\left( \Lambda -\frac 12\sigma _t\sigma \right) \left( \chi +2y\delta
_y\chi \right) +i\frac{r_t\sigma ^2}{\Omega _{in}}\delta _\tau \chi = 
$$

\begin{equation}
\label{eq4.6}
=-\frac 12\left\{ \delta _y^2-\sigma ^2\left[ 2\dot{\Lambda}-4\sigma
_t\sigma ^{-1}\Lambda +4\sigma ^{-2}\Lambda ^2+\sigma ^2\Omega ^2\left(
t;\left\{ W\right\} \right) \right] y^2\right\} \chi ,  
\end{equation}

where the functionals $\sigma \left( t;\left\{ W\right\} \right) $, $r\left(
t;\left\{ W\right\} \right) $ and $\Lambda \left( t;\left\{ W\right\}
\right) $ are still unknown. For their determination we shall suppose that
the following relations were observed:

\begin{equation}  
\label{eq4.7}
r_t\left( t;\left\{ W\right\} \right) =\frac{\Omega _{in}}{\sigma ^2\left(
t;\left\{ W\right\} \right) },
\end{equation}

\begin{equation}  
\label{eq4.8}
\Lambda \left( t;\left\{ W\right\} \right) =\frac 12\sigma _t\left(
t;\left\{ W\right\} \right) \sigma \left( t;\left\{ W\right\} \right) ,
\end{equation}

\begin{equation}  
\label{eq4.9}
2\dot \Lambda -4\sigma _t\sigma ^{-1}\Lambda +4\sigma ^{-2}\Lambda ^2+\sigma
^2\Omega ^2\left( t;\left\{ W\right\} \right) =\frac 12\Omega _{in}^2.
\end{equation}

It is easy to see that the system of equations (\ref{eq4.7})-(\ref{eq4.9}),
taking into account (\ref{eq3.20}) and relations (\ref{eq3.24})-(\ref{eq3.25}),
is equivalent to the model equation (\ref{eq3.1}) and, thereby to
Langevin-type nonlinear SDE (\ref{eq3.13}) with the complex initial
condition (\ref{eq3.15}). Using the expressions (\ref{eq4.7})-(\ref{eq4.9})
one can finally obtain from (\ref{eq4.6}):

\begin{equation}
\label{eq4.10}
i\delta _\tau \chi \left( y,\tau \right) =\frac 12\left( \delta _y^2+\Omega
_{in}^2y^2\right) \chi \left( y,\tau \right) .  
\end{equation}

The equation (\ref{eq4.10}) describes an autonomous system of the quantum
harmonic oscillator on the stochastic space-time continuum. The solution of
equation (\ref{eq4.10}) is obtained by conventional methods
and has the form

\begin{equation}  
\label{eq4.11}
\chi \left( y,\tau \right) =\left[ \frac{\left( \left. \Omega _{in}\right/
\pi \right) ^{1/2}}{2^nn!}\right] ^{1/2}\exp \left\{ -i\left( n+\frac
12\right) \Omega _{in}\tau -\frac 12\Omega _{in}y^2\right\} H_n\left( \sqrt{%
\Omega _{in}}y\right) .
\end{equation}

Combining (\ref{eq4.4}) and (\ref{eq4.11}) for the wave functional, evolving
from the initial pure state (\ref{eq2.7}), it is easy to find the expression
(\ref{eq4.1}). Taking into account (\ref{eq3.24})-(\ref{eq3.25}) one can find
the relation between the stochastic time $\tau \left( t;\left\{ W\right\}
\right) $ and the apparent time, the natural parameter $t$,

\begin{equation}  
\label{eq4.12}
\tau =\stackrel{t}{\mathrel{\mathop{\int }\limits_{-\infty }}}\frac{%
dt^{\prime }}{\sigma ^2\left( t^{\prime };\left\{ W^{\prime }\right\}
\right) }.
\end{equation}

Now take note of the very important property of the wave functional

\begin{equation}  
\label{eq4.13}
\left\langle \overline{\Psi _{stc}^{\left( +\right) }\left( m|x,t;\left\{
\xi \right\} \right) }\Psi _{stc}^{\left( +\right) }\left( n|x,t;\left\{ \xi
\right\} \right) \right\rangle _x=\delta _{mn},\qquad \left\langle \ldots
\right\rangle =\stackrel{+\infty }{\mathrel{\mathop{\int }\limits_{-\infty }}%
}dx,
\end{equation}

pointing to the fact that the basis formed in the space $L_2\left(
R^1\otimes R_{\left\{ \xi \right\} }\right) $ is orthonormal.

{\it The theorem is proved.}


\section{Derivation of Fokker-Plank equation for conditional probability
$P\left( {\bf \Phi },\lowercase{t}|{\bf \Phi }^{\prime },\lowercase{t}^{\prime
}\right) $}

Let us consider the functional of the form:

\begin{equation}  
\label{eq5.1}
P\left( {\bf \Phi },t|{\bf \Phi }^{\prime },t^{\prime }\right) =\left\langle
\delta \left[ {\bf \Phi }\left( t\right) -{\bf \Phi }\left( t^{\prime
}\right) \right] \right\rangle ,
\end{equation}

where ${\bf \Phi }\left( t\right) \equiv {\bf \Phi }\left( t;\left\{
W\right\} \right) $ is the solution of SDE (\ref{eq3.19}). After the
differentiation of functional (\ref{eq5.1}) over the time and use of (\ref
{eq3.19}) one can obtain:

$$
\partial _tP\left( {\bf \Phi },t|{\bf \Phi }^{\prime },t^{\prime }\right)
=-\partial _{{\bf \Phi }}\left\langle \stackrel{.}{\bf \Phi }\delta \left[ 
{\bf \Phi }\left( t\right) -{\bf \Phi }\left( t^{\prime }\right) \right]
\right\rangle = 
$$

\begin{equation}  
\label{eq5.2}
=\partial _{{\bf \Phi }}\left\{ {\bf K}\left( t;\left\{ {\bf \Phi }\right\}
\right) P\left( {\bf \Phi },t|{\bf \Phi }^{\prime },t^{\prime }\right)
+\left\langle {\bf F}\left( t;\left\{ W\right\} \right) \delta \left[ {\bf %
\Phi }\left( t\right) -{\bf \Phi }\left( t^{\prime }\right) \right]
\right\rangle \right\} .
\end{equation}

To obtain the equation for conditional probability (\ref{eq5.1}) in an
explicit form one has to specify the stochastic frequency ${\bf F}\left(
t;\left\{ W\right\} \right) $. Consider the most common case when the
stochastic component (\ref{eq3.20}) is the Gaussian random function $F\left(
t;\left\{ W\right\} \right)={W}=F\left( t\right) $. This assumption implies
that one can completely define $F\left( t\right) $ with the help of
correlation function that is a white noise type correlator in case when $%
F\left( t\right) $ changes faster than the solution $\theta \left( t\right) $

\begin{equation}  
\label{eq5.3}
\left\langle F\left( t\right) F\left( t^{\prime }\right) \right\rangle
=2\varepsilon \delta \left( t-t^{\prime }\right) ,\qquad \left\langle
F\left( t\right) \right\rangle =0,\quad \varepsilon >0.
\end{equation}

Now, using the Vick's theorem (see \cite{Gardiner})

\begin{equation}  
\label{eq5.4}
\left\langle F\left( t\right) N\left( t;\left\{ F\right\} \right)
\right\rangle =2\left\langle \frac{\delta N}{\delta F}\right\rangle
_{\left\{ F\right\} },
\end{equation}

where $N\left( t;\left\{ F\right\} \right) $ is an arbitrary functional of $%
F\left( t\right) $, one can write the following expression

\begin{equation}  
\label{eq5.5}
\left\langle {\bf F}\left( t;\left\{ W\right\} \right) \delta \left[ {\bf %
\Phi }\left( t\right) -{\bf \Phi }\left( t^{\prime }\right) \right]
\right\rangle =2\partial _\theta \left\langle \frac{\delta \theta \left(
t\right) }{\delta F\left( t\right) }\cdot \delta \left[ {\bf \Phi }\left(
t\right) -{\bf \Phi }\left( t^{\prime }\right) \right] \right\rangle .
\end{equation}

Due to the stochasticity of $\theta \left( t\right) $ the variational
derivative of $F\left( t\right) $ is $\varepsilon \cdot \mathop{\rm sgn}%
\left( t-t^{\prime }\right) +O\left( t-t^{\prime }\right) $. After common
regularization procedure (in the sense of Fourier decomposition) one can
find the value at $t=t^{\prime }$: \thinspace $\varepsilon \cdot \mathop{\rm %
sgn}\left( 0\right) =\frac 12\varepsilon $. Taking into account the
aforesaid we finally obtain the following Fokker-Plank equation for
conditional probability:

\begin{equation}
\label{eq5.6}
\partial _tP\left( {\bf \Phi },t|{\bf \Phi }^{\prime },t^{\prime }\right) =%
\stackrel{2}{\mathrel{\mathop{\sum }\limits_{i,j=1}}}\partial _{\Phi
_i}\left[ K_i\left( t;\left\{ {\bf \Phi }\right\} \right) +\varepsilon
_{ij}\partial _{\Phi _j}\right] P\left( {\bf \Phi },t|{\bf \Phi }^{\prime
},t^{\prime }\right) ,  
\end{equation}

where

$$
\varepsilon _{11}=\varepsilon ,\qquad \varepsilon _{12}=\varepsilon
_{21}=\varepsilon _{22}=0,\qquad \Phi _1=\theta ,\qquad \Phi _2=\varphi 
$$

\begin{equation}  
\label{eq5.7}
K_1\left( t;\left\{ {\bf \Phi }\right\} \right) =\left[ \theta ^2-\varphi
^2+U_0\left( t\right) \right] ,\qquad K_2\left( t;\left\{ {\bf \Phi }%
\right\} \right) =2\theta \varphi .
\end{equation}

Note, that the equation (\ref{eq5.6}) determines a diffusion process for
which ${\bf \Phi }\left( t\right) $ is continuous.

Let the conditional probability meet the boundary condition

\begin{equation}
\label{eq5.8}
P\left( {\bf \Phi },t|{\bf \Phi }^{\prime },t^{\prime }\right) =\delta
\left( {\bf \Phi }-{\bf \Phi }^{\prime }\right) , 
\end{equation}

then for small time intervals it is easy to determine the solution of
equation (\ref{eq5.6}):

$$
P\left( {\bf \Phi },t|{\bf \Phi }^{\prime },t^{\prime }\right) =\frac 1{2\pi 
\sqrt{\varepsilon \Delta t}}\times
$$

$$
\times \exp \left\{ -\frac 1{2\Delta t}\left[ {\bf \Phi 
} -{\bf \Phi }^{\prime }-{\bf K}\left( t;\left\{ {\bf \Phi }\right\} \right)
\Delta t\right] ^T\varepsilon _{ij}^{-1}\left( {\bf \Phi }-{\bf \Phi }%
^{\prime }-{\bf K}\left( t;\left\{ {\bf \Phi }\right\} \right) \Delta
t\right) \right\} = 
$$

\begin{equation}  
\label{eq5.9}
=\frac 1{2\pi \sqrt{\varepsilon \Delta t}}\exp \left\{ -\frac 1{2\varepsilon
\Delta t}\left[ \theta -\theta ^{\prime }-\left( \theta ^2-\varphi
^2+U_0\left( t\right) \right) \Delta t\right] ^2\right\} ,\qquad t=t^{\prime
}+\Delta t.
\end{equation}

Thus, one can state that the evolution of the system in the functional space 
$R_{\left\{ {\bf \Phi }\right\} }$ is characterized by a regular shift with
the velocity ${\bf K}\left( t;\left\{ {\bf \Phi }\right\} \right) $, against
the background of which the Gaussian fluctuations with diffusion matrix $%
\varepsilon _{ij}$ take place. As to the trajectory ${\bf \Phi }\left(
t\right) $ in the space $R_{\left\{ {\bf \Phi }\right\} }$, it is determined
by the formula (see \cite{Gardiner})

\begin{equation}  
\label{eq5.10}
{\bf \Phi }\left( t+\Delta t\right) ={\bf \Phi }\left( t\right) +{\bf K}%
\left( t;\left\{ {\bf \Phi }\right\} \right) \Delta t+{\bf F}\left( t\right)
\Delta t^{1/2}.
\end{equation}

The trajectory is seen from (\ref{eq5.10}) to be continuous everywhere, i.e. 
${\bf \Phi }\left( t+\Delta t\right) \mathrel{\mathop{\rightarrow }\limits%
_{\Delta t\rightarrow 0}}{\bf \Phi }\left( t\right) $, but is
undifferentiable everywhere owing to the presence of term $\sim \Delta
t^{1/2}$. If we write the time interval in the form $\Delta t=t/N$, where $%
N\rightarrow \infty $, then one can interpret the expression (\ref{eq5.9})
as the probability of transition from ${\bf \Phi }_k={\bf \Phi }\left(
t^{\prime }\right) $ to ${\bf \Phi }_{k+1}={\bf \Phi }\left( t\right) $
during the time $\Delta t$ in the model of Brownian motion.


\section{Solution of Fokker-Plank equation for the distribution of $\theta $
coordinate in the limit $\lowercase{t}\rightarrow +\infty $}

First, let us consider the equation (\ref{eq3.18}). Taking into account (\ref
{eq3.16}) as well as the initial condition (\ref{eq3.15}) one can have the
solution of equation (\ref{eq3.18}) in the explicit form:

\begin{equation}  
\label{eq6.1}
\varphi =\Omega _{in}\exp \left( -2\stackrel{t}{\mathrel{\mathop{\int
}\limits_{-\infty }}}\theta \left( t^{\prime }\right) dt^{\prime }\right)
\end{equation}

Using (\ref{eq6.1}) one can reduce the system of SDE (\ref{eq3.17})-(\ref
{eq3.18}) to one nonlinear and nonlocal SDE of the form:

\begin{equation}  
\label{eq6.2}
\dot \theta +\theta ^2-\Omega _{in}^2\exp \left( -4\stackrel{t}{%
\mathrel{\mathop{\int }\limits_{-\infty }}}\theta \left( t^{\prime }\right)
dt^{\prime }\right) +U_0\left( t\right) +F\left( t\right) =0.
\end{equation}

Using SDE (\ref{eq6.2}) one can obtain the equation for conditional
probability $Q\left( \theta ,t\right) =P\left( \theta ,t|0,0\right) $ in the
sense of distribution of coordinate $\theta $ at the time $t$:

\begin{equation}
\label{eq6.3}
\partial _tQ\left( \theta ,t\right) +\partial _\theta J\left( \theta ,%
\stackrel{t}{\mathrel{\mathop{\int }\limits_{-\infty }}}\theta \left(
t^{\prime }\right) dt^{\prime };t\right) =0  
\end{equation}

with the flow of probability

\begin{equation}  
\label{eq6.4}
J\left( \theta ,\stackrel{t}{\mathrel{\mathop{\int }\limits_{-\infty }}}%
\theta \left( t^{\prime }\right) dt^{\prime };t\right) =-\left[ \theta
^2-\Omega _{in}^2\exp \left( -4\stackrel{t}{\mathrel{\mathop{\int
}\limits_{-\infty }}}\theta \left( t^{\prime }\right) dt^{\prime }\right)
+U_0\left( t\right) \right] Q-\varepsilon \partial _\theta Q
\end{equation}

In view of the fact that in the equation (\ref{eq6.3}) the factor of the shift
$\left[ \theta ^2-\Omega _{in}^2 \exp \left( -4\times \right. \right. $
$\left. \left. \times \stackrel{t}{\mathrel{\mathop{\int
}\limits_{-\infty }}}\theta \left( t^{\prime }\right) dt^{\prime }\right)
+U_0\left( t\right) \right] $ tends to infinity at the limits
$\theta \rightarrow \mp \infty $, a symmetric non-vanishing flows of
probability $J\left( -\infty ,+\infty ;t\right) =J\left( +\infty ,-\infty
;t\right) \ne 0 $ there arise on the borders. For solution of equation
(\ref{eq6.3})-(\ref{eq6.4}) one has to require the natural
initial and boundary conditions

\begin{equation}
\label{eq6.5}
\stackunder{t\rightarrow t_c}{\lim }Q\left( \theta ;t\right) =\delta \left(
\theta \right) ,  
\end{equation}

\begin{equation}  
\label{eq6.6}
\mathrel{\mathop{\lim }\limits_{\left| \theta \right| \rightarrow \infty }}%
Q\left( \theta ;t\right) =0.
\end{equation}

The investigation of equation (\ref{eq6.3})-(\ref{eq6.4}) with initial and
boundary conditions for arbitrary $t$ is a very difficult
problem. However, below we shall see that for the construction of averaged
matrix in the model of roaming QRHO it is important to know the distribution
of $\theta $ coordinate when $t\rightarrow +\infty $.

Turning again to the solution of (\ref{eq6.1}) note, that in the limit of
large time intervals the self-averaging of trajectory $\theta \left( t\right) $
takes place by virtue of the ergotic hypothesis, so that

\begin{equation}  
\label{eq6.7}
\mathrel{\mathop{\lim }\limits_{t\rightarrow +\infty }}\varphi \left(
t\right) =\alpha \Omega _{in},\qquad \alpha =\exp \left( -2\stackrel{+\infty 
}{\mathrel{\mathop{\int }\limits_{-\infty }}}\theta \left( t^{\prime
}\right) dt^{\prime }\right) .
\end{equation}

We should like to remind, that by virtue of the same ergodic hypothesis the
constant $\alpha $ can be calculated by averaging over the ensemble.

Now taking into account that for $t\rightarrow +\infty $ the density of
flow reaches its limiting value

\begin{equation}  
\label{eq6.8}
J_{0f}=-\mathrel{\mathop{\lim }\limits_{t\rightarrow +\infty }}J\left( \theta ,%
\stackrel{t}{\mathrel{\mathop{\int }\limits_{-\infty }}}\theta \left(
t^{\prime }\right) dt^{\prime };t\right) ,
\end{equation}

we obtain the equation for the probability of distribution of a stationary
process,

\begin{equation}  
\label{eq6.9}
J_{0f}=\left[ \theta ^2-\alpha ^2\Omega _{in}^2+U_{+}\right] Q_s-\varepsilon
d_\theta Q_s,\qquad d_\theta =\left. d\right/ d\theta ,
\end{equation}

$$
U_{+}=\mathrel{\mathop{\lim }\limits_{t\rightarrow +\infty }}U_0\left(
t\right) . 
$$

In Eq.(\ref{eq6.9}) the probability distribution for a stationary process is
denoted by $Q_s$ that is obtained from Eq.(\ref{eq6.9}) to be \cite{Gredeskul}

$$
Q_s\left( \varepsilon ,\lambda ,\gamma ;\theta \right) =\varepsilon
^{-1/3}\bar Q_s\left( \lambda ,\gamma ;\bar \theta \right) = 
$$

\begin{equation}  
\label{eq6.10}
=\varepsilon ^{-2/3}J_{0f}\exp \left( -\frac{\bar \theta ^3}3-\lambda \gamma
\bar \theta \right) \stackrel{\bar \theta }{\mathrel{\mathop{\int
}\limits_{-\infty }}}dz\exp \left( \left( \frac{z^3}3+\lambda \gamma
z\right) \right) ,
\end{equation}

where $\lambda =\left( \left. \Omega _{in}\right/ \varepsilon ^{1/3}\right)
^2,$ $\gamma =\left( \left. \Omega _{out}\right/ \Omega _{in}\right)
^2-\alpha ^2+\left. F_0\right/ \Omega _{in}^2$ and $\bar \theta =\left.
\theta \right/ \varepsilon ^{1/3}$. The constant $J_{0f}$ is calculated from
the condition of normalization of stationary distribution $Q_s$ to unity and
has the form

\begin{equation}  
\label{eq6.11}
J_{0f}^{-1}=\pi ^{1/2}\varepsilon ^{-1/3}\stackrel{\infty }{\mathrel{\mathop{%
\int }\limits_{0}}}dz\,z^{-1/2}\exp \left( \frac{z^3}{12}-\lambda \gamma
z\right) .
\end{equation}

One can obtain representation of $J_{0f}$ in terms of special functions. Passing
to Fourier components in equation (\ref{eq6.9}) we find \cite{Lax}:

\begin{equation}
\label{eq6.12}
\overline{J}_{0f}^{-1}=\left[ Ai^2\left( -\lambda \gamma \right) +Bi^2\left(
-\lambda \gamma \right) \right] ,
\end{equation}

where $Ai\left( x\right) $ and $Bi\left( x\right) $ are linear independent 
solutions of Airy equation \cite{Abramovitz}

\begin{equation}
\label{eq6.13}
y^{^{\prime \prime }}-xy=0.
\end{equation}

Note, that the normalization constant $J_{0f}$ has one important feature,
namely, it defines the number of states with energy values $E$ from $-\infty $
up to $\Omega ^2_{out}$ per unit of distance. In the case when $\left(
E/\varepsilon^{2/3}\right) \gg 1$ and $E>0$ the expression (\ref{eq6.11})
is calculated asymptotically:

\begin{equation}
\label{eq6.14}
N_{\Sigma }=J_{0f}\approx \pi ^{-1}E^{1/2}\left[ 1+\frac 5{32}\frac{\varepsilon
^2}{E^3}+O\left( \frac{\varepsilon ^4}{E^6}\right) \right] .
\end{equation}

One can use expression (\ref{eq6.11}) for determining of states distribution
for concrete energy value. Taking into account the fact that bound states
are characterized by negative energy values and making in equation (\ref{eq6.11})
the formal substitution $\Omega ^2_{out}+F_0\rightarrow -E$ one can obtain
expression for the number of states with concrete localization energy:

\begin{equation}
\label{eq6.15}
N_E=\pi ^{-1}\varepsilon ^{1/2}\stackrel{\infty }{\stackunder{0}{\int }}%
dzz^{-1/2}\exp \left( -\frac{z^3}{12}+\frac E{\varepsilon ^{2/3}}z\right) .
\end{equation}

In the case when $\left( E/\varepsilon^{2/3}\right) \gg 1$ and $E>0$
the expression (\ref{eq6.15}) is calculated asymptotically and have the
following form:

\begin{equation}
\label{eq6.16}
N_E\approx \pi ^{-1}E^{1/2}\exp \left( -\frac 43\frac{E^{3/2}}\varepsilon
\right) \left[ 1+O\left( \frac \varepsilon {E^{3/2}}\right) \right] .
\end{equation}

It is easy to calculate now the states distribution for concrete value of
bound state energy:

\begin{equation}
\label{eq6.17}
P_E=\frac{N_E}{N_{\Sigma }}\approx \exp \left( -\frac 43\frac{E^{3/2}}\varepsilon
\right) \left[ 1+O\left( \frac \varepsilon {E^{3/2}}\right) \right] .
\end{equation}

It is worthwhile at the end of this section to pay attention to one
important property of the stationary distribution $Q_s$, that follows, in
particular, from numerical analyses. The fact is that for all $\gamma \in
\left( -\infty ,+\infty \right) $ the probability of positive values of the 
$\theta $ coordinate is higher than that for negative values
(see FIG.1).
It follows, hence, that $\alpha =0$ and $\varphi _s=0$ , where $\varphi _s$
is the imaginary part of complex coordinate $\Phi $ in the stationary process
limit.


\section{Calculation of the wave function of roaming QRHO}

Having at hand the expression for conditional probability (\ref{eq5.9}) one
can now pass to the averaging of the wave functional in the $R_{\left\{ \xi
\right\} }$ space, or, in other words, to the calculation of wave function
of roaming QRHO. Writing the wave functional (\ref{eq4.1}) in the moving
reference system (see (\ref{eq3.12})), and taking into account the formula (%
\ref{eq2.7}) for the average value of wave functional, we obtain

$$
\Psi _{br}^{\left( +\right) }\left( n|x,t\right) =\left\langle \Psi
_{stc}^{\left( +\right) }\left( n|x,t;\left\{ \xi \right\} \right)
\right\rangle _{\left\{ \xi \right\} }=\left\langle \tilde \Psi
_{stc}^{\left( +\right) }\left( n|x,t;\left\{ {\bf \Phi }\right\} \right)
\right\rangle _{\left\{ {\bf \Phi }\right\} }= 
$$

\begin{equation}
\label{eq7.1}
\frac 1\alpha \int D\mu \left\{ {\bf \Phi }\right\} \tilde{\Psi}%
_{stc}^{\left( +\right) }\left( n|x,t;\left\{ {\bf \Phi }\right\} \right) ,
\end{equation}

In this formula $D\mu \left\{{\bf \Phi } \right\} $ determines the total
Fokker-Plank measure of the functional space $R_{\left\{ {\bf \Phi }
\right\} }$ :

\begin{equation}
\label{eq7.2}
\begin{array}{c}
D\mu \left\{ {\bf \Phi }\right\} =d\mu \left\{ {\bf \Phi }_0\right\} \cdot
d\mu \left\{ {\bf \Phi }_t\right\} \cdot \mathrel{\mathop{\lim
}\limits_{N\rightarrow \infty }}\left[ \left( \frac 1{2\pi }\sqrt{\frac
N{\varepsilon t}}\right) ^N \times \right.
\\ \\
\left. \times \mathrel{\mathop{%
\stackrel{N}{\prod }}\limits_{k=0}}\exp \left\{ -\frac N{2\varepsilon
t}\left[ \theta -\theta ^{^{\prime }}-\left( \theta ^2+\varphi
^2+U_{0}\left( t\right) \right) \frac tN\right] ^2\right\} d\theta
_{k+1}d\varphi _{k+1}\right] ,
\end{array}
\end{equation}

where $\alpha ,$ $d\mu \left\{ {\bf \Phi }_0\right\} $ and $d\mu \left\{ 
{\bf \Phi }_t\right\} $ are determined with the help of formulae

\begin{equation}
\begin{array}{c}
\label{eq7.3}
\alpha \left( \varepsilon ,t\right) =\int D\mu \left\{ {\bf \Phi }\right\} ,
\\ \\
d\mu \left\{ {\bf \Phi }_0\right\} =\delta \left( \theta _0 \right)
\delta \left( \Omega _{in}-\varphi _0\right) d\theta _0d\varphi _0,
\\ \\
d\mu \left\{ {\bf \Phi }_t\right\} =P\left( {\bf \Phi },t\mid 0,0\right)
d\theta d\varphi .
\end{array}
\end{equation}

In equations (\ref{eq7.2})-(\ref{eq7.3}) $\alpha (\varepsilon ,t)$ denotes
the normalization factor for functional integral (\ref{eq7.1}) with total
Fokker-Plank measure and in the limit $t\rightarrow +\infty $ it acquires a
constant value that, in general, is different from the unity. The
integration over the measure $d\mu \left\{ {\bf \Phi }_0\right\} $ means the
averaging in the initial distribution of complex coordinate $\Phi $ at the
moment of time $t=-\infty $. As for the integration over the measure $d\mu
\left\{ {\bf \Phi }_t\right\} $ , it provides the averaging in the
distribution of complex coordinate $\Phi $ at an arbitrary instant $t$ . In
particular, one can show that in the limit $t\rightarrow +\infty $ it
assumes the following form:

\begin{equation}
\label{eq7.4}
d\mu \{{\bf \Phi }_\infty \}=d\mu \{{\bf \Phi }_s\}=Q_s(\varepsilon ,\lambda
,\gamma ;\theta )\delta (\varphi )d\theta d\varphi .  
\end{equation}

Now we shall try to study the $\left( out\right) $ asymptotical state of the
"oscillator+thermostat" system .

Taking into account the relation (\ref{eq2.5}) one can separate the
fluctuation process at the frequency (\ref{eq2.3}) in the span of time $%
(t_c,+\infty _{}),$ where $t_c$ is the zero time reference point in the $%
\left( out\right) $ channel, by giving it with the help of white noise
correlator with the diffusion constant $\varepsilon _{+}.$ It is evident,
hence, that the $\left( out\right) $ asymptotic state of the
"oscillator+thermostat" system will be also characterized by a complex
probabilistic process $\Psi _{out}\left( m|x,t;\left\{ \eta _{+}\right\}
\right) $ in the extended space $\Xi _{out}=R_{out}^1\otimes R_{\left\{ \eta
_{+}\right\} }$, where $\eta _{+}=\exp \left( i\Omega _{out}t+\stackrel{t}{%
\mathrel{\mathop{\int
}\limits_{t_c}}}\Phi _{+}\left( t^{\prime }\right) dt^{\prime }\right) $ is
the corresponding solution of SDE (\ref{eq3.1}). Proceeding with the same
reasoning one can construct the wave function of the $\left( out\right) $
asymptotic state $\Psi _{br}\left( m|x,t\right) $.

At the end of this section we should like to note that in the limit $%
\varepsilon \rightarrow 0$ the measure of the functional space $R_{\left\{
\xi \right\} }$ turns out to be of the Wiener Type, and the wave function of
roaming QRHO, $\Psi _{br}^{\left( +\right) }\left( n|x,t\right) $ steadily
transforms to the exact solution $\Psi ^{\left( +\right) }\left(
n|x,t\right) $ that describes the motion of regularly moving QRHO.


\section{The local stochastic matrix of transitions of roaming QRHO}

As was mentioned above, the roaming QRHO has two asymptotic states: $\Psi
_{in}\left( n|x,t\right) \in L_2\left( R_{in}^1\right) $ when $t\rightarrow
-\infty $ (see (\ref{eq2.7})) and respectively $\Psi _{out}\left(
m|x,t;\left\{ \eta _{+}\right\} \right) \in L_2\left( R_{out}^1\otimes
R_{(\eta _{+})}\right) $ when $t\rightarrow +\infty $. The goal of the
theory of scattering is to construct a unitary operator that will change one
of these states to the other \cite{Lax}.

Prior to that let us consider the wave functional $\Psi _{stc}^{\left(
-\right) }\left( m|x,t;\left\{ \eta \right\} \right) $ that is the solution
of SDE (\ref{eq2.1})-(\ref{eq2.2}) and passes to the asymptotic wave
functional $\Psi _{out}\left( m|x,t_{+};\left\{ \eta _{+}\right\} \right) $
when $t=t_{+}>t_c$, which in its turn passes to the wave function $\Psi
_{out}\left( m|x,t\right) $ when $t=t_c$. Since the sets of functionals $%
\Psi _{stc}^{\left( +\right) }\left( n|x,t;\left\{ \xi \right\} \right) $
and $\Psi _{stc}^{\left( -\right) }\left( m|x,t;\left\{ \eta \right\}
\right) $ form completely orthonormalized bases in $L_2$, one can write the
following decompositions:

\begin{equation}  
\label{eq8.1}
\Psi _{stc}^{\left( +\right) }\left( n|x,t;\left\{ \xi \right\} \right) =%
\mathrel{\mathop{\sum }\limits_{k}}S_{kn}\left( t;\left\{ \xi \right\}
|t^{\prime };\left\{ \eta ^{\prime }\right\} \right) \Psi _{stc}^{\left(
-\right) }\left( k|x,t^{\prime };\left\{ \eta ^{\prime }\right\} \right) ,
\end{equation}

\begin{equation}  
\label{eq8.2}
\Psi _{stc}^{\left( -\right) }\left( m|x,t;\left\{ \eta \right\} \right) =%
\mathrel{\mathop{\sum }\limits_{k}}S_{km}^{*}\left( t^{\prime };\left\{ \xi
^{\prime }\right\} |t;\left\{ \eta \right\} \right) \Psi _{stc}^{\left(
+\right) }\left( k|x,t^{\prime };\left\{ \xi ^{\prime }\right\} \right) ,
\end{equation}

where the coefficients $S_{kn}$ and $S_{km}^{*}$ are some random local
evolutionary operators. Taking into account the orthogonality relations (\ref
{eq4.13}), one can obtain from (\ref{eq8.1})

\begin{equation}
\label{eq8.3}
S_{nm}\left( t;\left\{ \xi \right\} |t^{\prime };\left\{ \eta ^{\prime
}\right\} \right) =\left\langle \Psi _{stc}^{\left( +\right) }\left(
n|x,t;\left\{ \xi \right\} \right) \stackrel{}{\overline{\Psi _{stc}^{\left(
-\right) }\left( m|x,t^{\prime };\left\{ \eta ^{\prime }\right\} \right) }}%
\right\rangle _x.  
\end{equation}

One can easily find from expressions (\ref{eq8.1}) and (\ref{eq8.2}), with
due regard for (\ref{eq4.13}) and (\ref{eq8.3}),

\begin{equation}
\label{eq8.4}
\mathrel{\mathop{\sum }\limits_{k}}S_{kn}\left( t;\left\{ \xi \right\}
|t^{\prime };\left\{ \eta ^{\prime }\right\} \right) S_{km}^{*}\left(
t;\left\{ \xi \right\} |t^{\prime };\left\{ \eta ^{\prime }\right\} \right)
=\delta _{nm,}  
\end{equation}

that is basically a generalization of the unitarity condition for the
stochastic matrix $S^{stc}$, the matrix elements of which are determined by
means of the formula (\ref{eq8.3}).

Note that the unitarity of the stochastic matrix $S^{stc}\left(
S^{stc}\right) ^{+}=\mathop{\rm I}$ is a consequence of conservation of laws
in the closed system "oscillator+thermostat", that is defined in the
extended space $\Xi $.

It is convenient to carry out explicit calculations of the elements of
stochastic matrix (\ref{eq8.3}) by means of the method of generating
functionals. Let us consider the sum

\begin{equation}
\label{eq8.5}
\Psi _{stc}^{\left( +\right) }\left( z|x,t;\left\{ \xi \right\} \right) =%
\stackrel{\infty }{\mathrel{\mathop{\sum }\limits_{n=0}}}\frac{z^n}{\sqrt{n!}%
}\Psi _{stc}^{\left( +\right) }\left( n|x,t;\left\{ \xi \right\} \right) ,
\end{equation}

where $z$ is an auxiliary function. If we substitute the expression for wave
functional (\ref{eq4.1}) in (\ref{eq8.5}) and make an appropriate summation
(see \cite{Abramovitz}), then we shall have

\begin{equation}  
\label{eq8.6}
\Psi _{stc}^{\left( +\right) }\left( z|x,t;\left\{ \xi \right\} \right)
=\left( \frac{\Omega _{in}}\pi \right) ^{\frac 14}\exp \left\{ -\frac
12\left( ax^2-2bx+c\right) \right\} ,
\end{equation}

where the coefficients $a$, $b$ and $c$ are

\begin{equation}  
\label{eq8.7}
\begin{array}{c}
a\left( t;\left\{ \xi \right\} \right) =-i \frac{\dot \xi \left( t;\left\{
W\right\} \right) }{\xi \left( t;\left\{ W\right\} \right) }, \\ 
\\ 
b\left( t;\left\{ \xi \right\} \right) = \frac{\sqrt{2\Omega _{in}}z}{\xi
\left( t;\left\{ W\right\} \right) }, \\ 
\\ 
c\left( t;\left\{ \xi \right\} \right) =z^2\exp \left[ -i2r\left( t;\left\{
W\right\} \right) \right] .
\end{array}
\end{equation}

As it is seen from (\ref{eq8.6})-(\ref{eq8.7}), the generating functional in $x$
coordinate is a stochastic Gaussian packet that in the limit $t\rightarrow
-\infty $ goes into an ordinary Gaussian packet:

$$
\Psi _{stc}^{\left( +\right) }\left( z|x,t;\left\{ \xi \right\} \right) %
\mathrel{\mathop{\rightarrow }\limits_{t\rightarrow -\infty }}\Psi
_{in}\left( z|x,t\right) = \left( \frac{\Omega _{in}}\pi \right) ^{\frac 14}
\times
$$

\begin{equation}  
\label{eq8.8}
=\times \exp \left\{ -\frac
12\left( \Omega _{in}x^2-2\sqrt{\Omega _{in}}zx\exp \left( -i\Omega
_{in}t\right) +z^2\exp \left( -i2\Omega _{in}t\right) +i\Omega _{in}t\right)
\right\} .
\end{equation}

As for the generating functional of the $\left( out\right) $ state, it is
easily obtained from (\ref{eq8.6})-(\ref{eq8.7}) by making simple formal
substitutions $z\rightarrow z_{+},$ $\Omega _{in}\rightarrow \Omega _{out},$ 
$\xi \rightarrow \eta _{+}$ and $t\rightarrow t_{+}$:

\begin{equation}  
\label{eq8.9}
\Psi _{out}\left( z_{+}|x,t_{+};\left\{ \eta _{+}\right\} \right) =\left( 
\frac{\Omega _{in}}\pi \right) ^{\frac 14}\eta _{+}^{-1/2}\exp \left\{
-\frac 12\left( a_{+}x^2-2b_{+}x+c_{+}\right) \right\} ,
\end{equation}

where the following notations are made:

\begin{equation}  
\label{eq8.10}
\begin{array}{c}
a_{+}\left( t_{+};\left\{ \eta _{+}\right\} \right) =-i \frac{\dot \eta
\left( t_{+};\left\{ W_{+}\right\} \right) }{\eta _{+}\left( t_{+};\left\{
W_{+}\right\} \right) }, \\ 
\\ 
b_{+}\left( t_{+},z_{+};\left\{ W_{+}\right\} \right) = \frac{\sqrt{2\Omega _{out}}%
z_{+}}{\eta _{+}\left( t_{+};\left\{ W_{+}\right\} \right) }, \\ 
\\ 
c_{+}\left( t_{+};\left\{ \eta _{+}\right\} \right) =z_{+}^2\exp \left[
-i2r_{+}\left( t_{+};\left\{ W_{+}\right\} \right) \right] , \\ 
\\ 
r_{+}\left( t_{+};\left\{ W_{+}\right\} \right) =\Omega _{out}\stackrel{t_{+}%
}{\mathrel{\mathop{\int }\limits_{t_c}}}\frac{dt^{\prime }}{\eta _{+}\left(
t^{\prime };\left\{ W_{+}^{\prime }\right\} \right) } .
\end{array}
\end{equation}

Now consider the following integral:

\begin{equation}  
\label{eq8.11}
I\left( z,t;\left\{ \xi \right\} |z_{+}^{*},t_{+};\left\{ \eta
_{+}^{*}\right\} \right) =\left\langle \Psi _{stc}^{\left( +\right) }\left(
z|x,t;\left\{ \xi \right\} \right) \overline{\Psi _{out}\left(
z_{+}|x,t_{+};\left\{ \eta _{+}\right\} \right) }\right\rangle _x.
\end{equation}

Making appropriate substitutions for the generating functionals from (\ref
{eq8.6})-(\ref{eq8.7}) and (\ref{eq8.8})-(\ref{eq8.9}) and integrating over
the coordinate $x$ we obtain

\begin{equation}  
\label{eq8.12}
I\left( z,t;\left\{ \xi \right\} |z_{+}^{*},t_{+};\left\{ \eta
_{+}^{*}\right\} \right) =\left( \Omega _{in}\Omega _{out}\right)
^{1/4}\left[ \frac 2{A\xi \eta _{+}^{*}}\right] ^{1/2}\exp \left\{ -\frac
12\left( C-\frac{B^2}A\right) \right\} ,
\end{equation}

with the following notations:

\begin{equation}  
\label{eq8.13}
\begin{array}{c}
A\left( t;\left\{ \xi \right\} |t_{+};\left\{ \eta _{+}^{*}\right\} \right)
=-i \frac{\dot \xi }\xi +i\frac{\dot \eta _{+}^{*}}{\eta _{+}^{*}}, \\ 
\\ 
B\left( t,z;\left\{ \xi \right\} |t_{+},z_{+};\left\{ \eta _{+}^{*}\right\} \right)
= \frac{\sqrt{\Omega _{in}}z}\xi +\frac{\sqrt{\Omega _{out}}z_{+}^{*}}{\eta
_{+}^{*}}, \\ 
\\ 
C\left( t;\left\{ \xi \right\} |t_{+};\left\{ \eta _{+}^{*}\right\} \right)
=\exp \left( -i2r\right) z^2+\exp \left( i2r_{+}\right) \left(
z_{+}^{*}\right) ^2.
\end{array}
\end{equation}

It is easy to show that $I\left( z,t;\left\{ \xi \right\}
|z_{+}^{*},t_{+};\left\{ \eta _{+}^{*}\right\} \right) $ is a generating
functional for stochastic matrix elements:

\begin{equation}  
\label{eq8.14}
I\left( z,t;\left\{ \xi \right\} |z_{+}^{*},t_{+};\left\{ \eta
_{+}^{*}\right\} \right) =\stackrel{\infty }{\mathrel{\mathop{\sum
}\limits_{n,m=0}}}\frac{z^nz_{+}^m}{\sqrt{n!m!}}S_{nm}\left( t;\left\{ \xi
\right\} |t_{+};\left\{ \eta _{+}^{*}\right\} \right) .
\end{equation}

Making expansion in the left hand side of expression (\ref{eq8.14}) in the
Taylor series in $z$ and $z_{+}^{*}$ we find

\begin{equation}  
\label{eq8.15}
S_{nm}\left( t;\left\{ \xi \right\} |t_{+};\left\{ \eta _{+}^{*}\right\}
\right) =\frac 1{\sqrt{n!m!}}\left[ \partial _z^n\partial
_{z_{+}^{*}}^mI\left( z,t;\left\{ \xi \right\} |z_{+}^{*},t_{+};\left\{ \eta
_{+}^{*}\right\} \right) \right] _{z=z_{+}^{*}=0}.
\end{equation}

To calculate several first elements of the local stochastic transition
matrix one can substitute (\ref{eq8.12}) to (\ref{eq8.15}):

\begin{equation}  
\label{eq8.16}
\begin{array}{c}
S_{00}\left( t;\left\{ \xi \right\} |t_{+};\left\{ \eta _{+}^{*}\right\}
\right) =\left( 4\Omega _{in}\Omega _{out}\right) ^{1/4}\exp \left( \frac{%
i\pi }4\right) \left\{ \xi \eta _{+}^{*}\left( \dot \xi \xi ^{-1}-\dot \eta
_{+}^{*}\left( \eta _{+}^{*}\right) ^{-1}\right) \right\} ^{-1/2}, \\ 
\\ 
S_{11}\left( t;\left\{ \xi \right\} |t_{+};\left\{ \eta _{+}^{*}\right\}
\right) =\left[ S_{00}\left( t;\left\{ \xi \right\} |t_{+};\left\{ \eta
_{+}^{*}\right\} \right) \right] ^3, \\ 
\\ 
S_{20}\left( t;\left\{ \xi \right\} |t_{+};\left\{ \eta _{+}^{*}\right\}
\right) =S_{00}\left[ -\exp \left( -i2r\right) +\left( \frac{\Omega _{in}}{%
\Omega _{out}}\right) ^{1/2}\xi ^{-1}\eta _{+}^{*}S_{00}^2\right] , \\ 
\\ 
S_{02}\left( t;\left\{ \xi \right\} |t_{+};\left\{ \eta _{+}^{*}\right\}
\right) =S_{00}\left[ -\exp \left( i2r_{+}\right) +\left( \frac{\Omega _{out}%
}{\Omega _{in}}\right) ^{1/2}\xi \left( \eta _{+}^{*}\right)
^{-1}S_{00}^2\right] .
\end{array}
\end{equation}

One can see from expressions (\ref{eq8.16}) that in analogy to the case of
regular QRHO the transitions will occur only between asymptotic states of
similar parity irrespective of the value of diffusion constant $\varepsilon $%
. But in contrast to the regular case, it is characteristic that the
symmetry in local matrix elements (\ref{eq8.16}) with respect to the
transposition of quantum numbers of initial $n$ and final $m$ channels is
violated.


\section{The averaged transition matrix of roaming QRHO. The probability of
"vacuum-vacuum" transition}

The ultimate objective of our study is the calculation of transition
probabilities of roaming QRHO. This aim in view it is necessary first to
make the averaging of local stochastic elements (\ref{eq8.16}).

{\bf Definition 1.} {\it The expression }

{\it 
\begin{equation}  
\label{eq9.1}
S_{nm}^{br}=\mathrel{\mathop{\lim }\limits_{ \begin{array}{c} t\rightarrow
+\infty \\ t_{+}\rightarrow +\infty \end{array} }}\left\langle \left\langle
\left\langle \left\langle \tilde S_{nm}\left( t;\left\{ {\bf \Phi }+{\bf f}%
\right\} |t_{+};\left\{ {\bf \Phi }_{+}+{\bf f}_{+}\right\} \right)
\right\rangle _{{\bf f}}\right\rangle _{{\bf f}_{+}}\right\rangle _{\left\{ 
{\bf \Phi }\right\} }\right\rangle _{\left\{ {\bf \Phi }_{+}\right\} }
\end{equation}
}

{\it is termed the averaged (generalized) transition matrix of roaming QRHO},

where the local $\tilde{S}_{nm}\left( t;\left\{ {\bf \Phi }+{\bf f}\right\}
|t_{+};\left\{ {\bf \Phi }_{+}+{\bf f}_{+}\right\} \right) $ matrix
element in (\ref{eq9.1}) is obtained from (\ref{eq8.16}) by writing the
latter in an arbitrary movable reference frame with the help of
transformation (\ref{eq3.12}). Remember that the first bracket in (\ref
{eq9.1}) means a functional integration over the measure

\begin{equation}
\label{eq9.2}
\begin{array}{c}
D{\bf f}\left( t\right) =\mathrel{\mathop{\lim }\limits_{N\rightarrow \infty
}}\mathrel{\mathop{\stackrel{N}{\prod }}\limits_{k=0}}\delta \left( {\bf f}%
\left( t_k\right) -{\bf f}_0\left( t_k\right) \right) d{\bf f}\left(
t_k\right) ,\qquad t_k=\frac tN, \\ 
\\ 
{\bf f}_0\left( t\right) =\left\{ \mathop{\rm Re}\left[ \frac{\dot{\xi}%
_{0t}\left( t\right) }{\xi _0\left( t\right) }\right] ;\mathop{\rm Im}\left[ 
\frac{\dot{\xi}_{0t}\left( t\right) }{\xi _0\left( t\right) }\right]
\right\} .
\end{array}
\end{equation}

The second bracket respectively means the integration over the measure

\begin{equation}
\label{eq9.3}
\begin{array}{c}
D{\bf f}_{+}\left( t\right) =Df_{+}\left( t\right) =\mathrel{\mathop{\lim
}\limits_{N\rightarrow \infty }}\mathrel{\mathop{\stackrel{N}{\prod
}}\limits_{k=0}}\delta \left( f_{+}\left( t_k\right) -f_{0+}\left(
t_k\right) \right) df_{+}\left( t_k\right) ,\qquad t_k=\frac tN, \\ 
\\ 
{\bf f}_{0+}=f_{0+}=\left\{ 0;\Omega _{out}\right\} .
\end{array}
\end{equation}

In (\ref{eq9.1}) the third and forth brackets respectively stand for
integration in $R_{\left\{ {\bf \Phi }\right\} }$ and $R_{\left\{ {\bf \Phi }%
_{+}\right\} }$ spaces.

Note that the integration over variables ${\bf f}\left( t\right) $ and $%
f_{+}\left( t\right) $ gives the average value of transition matrix in the
steady (laboratory) frame of reference. It is also noteworthy that, as it
follows from expressions (\ref{eq8.3})-(\ref{eq8.4}), the generalized
transition matrix $S^{br}$ is usually not a unitary matrix, i.e., $%
S^{br}\left( S^{br}\right) ^{+}\ne \mathop{\rm I}$. Nevertheless, there are
some limiting cases when one can simplify both the measures of functional
spaces $R_{\left\{ {\bf \Phi }\right\} }$ and $R_{\left\{ {\bf \Phi }%
_{+}\right\} }$ (by making these of the Wiener type), and the corresponding
wave functionals. As a result of these simplifications, the contribution to
the expression (\ref{eq9.1}) is made only by the integrals over the final
distributions of coordinates $\theta $ and $\theta _{+}$ and that, in its
turn, makes unitary the generalized transition matrix $S_W^{br}\left(
S_W^{br}\right) ^{+}=\mathop{\rm I}$, where the index $w$ means
simplification of the matrix in the above sense.

As an illustration of the proposed approach, consider now the probability of
"vacuum-vacuum" transition in the case when the spaces $R_{\left\{ {\bf \Phi 
}\right\} }$ and $R_{\left\{ {\bf \Phi }_{+}\right\} }$ have the Wiener-type
measures. After the substitution of the expression for $S_{00}\left(
t;\left\{ {\bf \Phi }\right\} |t_{+};\left\{ {\bf \Phi }_{+}\right\} \right) 
$ to (\ref{eq9.1}) and simple integration we find

\begin{equation}  
\label{eq9_4}
S_{00}^{br}\left( \lambda ,\lambda _{+};\rho \right) =\left( 1-\rho \right)
^{1/4}\left\{ I_1\left( \lambda ,\lambda _{+};\rho \right) -iI_2\left(
\lambda ,\lambda _{+};\rho \right) \right\} ,
\end{equation}

where the following notations were made:

\begin{equation}  
\label{eq9.5}
I_{1,2}\left( \lambda ,\lambda _{+};\rho \right) =\stackrel{+\infty }{%
\mathrel{\mathop{\int }\limits_{-\infty }}}d\bar \theta \bar Q_s\left(
\lambda ,\gamma ;\bar \theta \right) \stackrel{+\infty }{\mathrel{\mathop{%
\int }\limits_{-\infty }}}d\bar \theta _{+}\bar Q_s\left( \lambda
_{+},\gamma ;\bar \theta _{+}\right) \left( \frac{a\pm 1}{2a^2}\right)
^{1/2},
\end{equation}

\begin{equation}  
\label{eq9.6}
a\left( \lambda ;\bar \theta |\lambda _{+};\bar \theta _{+}\right) =\left[
1+\frac 1{\lambda \gamma }\left( \bar \theta -\sqrt{\frac \lambda {\lambda
_{+}}}\bar \theta _{+}\right) ^2\right] ^{1/2}.
\end{equation}

In above formulae $\rho $ stands for the reflection coefficient
from a barrier in the one-dimensional problem of quantum
mechanics with momentum $K\left( x\right) =\Omega _0\left( x\right) $, where 
$t$ was replaced by $x$ (see \cite{Landafshitz}). As for the function $\gamma
(\rho)$, for the frequency model in the form of step barrier

\begin{equation}
\label{eq9.7}
\Omega _0(t)=\Omega _{in}+(\Omega _{out}-\Omega _{in})\limfunc{sign}(t)  
\end{equation}

it has the form

\begin{equation}  
\label{eq9.8}
\gamma \left( \rho \right) =\left( \frac{\Omega _{out}}{\Omega _{in}}\right)
^2=\left( \frac{1+\rho ^{1/2}}{1-\rho ^{1/2}}\right) ^2.
\end{equation}

Now one can write the final expression for the amplitude of "vacuum-vacuum"
transition probability

\begin{equation}
\label{eq9.9}
\Delta _{0\rightarrow 0}^{br}\left( \lambda ,\lambda _{+};\rho \right) =%
\sqrt{1-\rho }\left\{ I_1^2\left( \lambda ,\lambda _{+};\rho \right)
+I_2^2\left( \lambda ,\lambda _{+};\rho \right) \right\} ,  
\end{equation}

with due regard for the modification of the nature of fluctuation process in
the "oscillator-thermostat" system.

In case when $\varepsilon _{+}\rightarrow 0$, i.e., the final state of the
system is described by the wave function, the expression (\ref{eq9.8}) is
strongly simplified

\begin{equation}
\label{eq9.10}
\overline{\Delta }_{0\rightarrow 0}^{br}\left( \lambda ;\rho \right) =\sqrt{%
1-\rho }\left\{ \bar{I}_1^2\left( \lambda ;\rho \right) +\bar{I}_2^2\left(
\lambda ;\rho \right) \right\} ,  
\end{equation}

\begin{equation}  
\label{eq9.11}
\bar I_{1,2}\left( \lambda ;\rho \right) =\stackrel{+\infty }{%
\mathrel{\mathop{\int }\limits_{-\infty }}}d\bar \theta \left( \frac{\bar a+1%
}{2\bar a^2}\right) ^{1/2}\bar Q_s\left( \lambda ,\gamma ;\bar \theta
\right) ,
\end{equation}

\begin{equation}  
\label{eq9.12}
\bar a\left( \lambda ,\gamma ;\bar \theta \right) =\left( 1+\frac{\bar
\theta ^2}{\lambda \gamma }\right) ^{1/2}.
\end{equation}

The numerical calculations of "vacuum-vacuum" transition probability by the
formulae (\ref{eq9.10})-(\ref{eq9.12}) show its nonmonotonic behaviour
(see FIG.2).


\section{Calculation of average values of dynamical variables. Thermodynamics
of the nonrelativistic vacuum in asymptotic subspace $R_{\lowercase{as}}^1$}

It is well known that the main object of quantum statistical mechanics is
the density matrix that can be written in the nonstationary representation
as (see \cite{Zubarev})

\begin{equation}
\label{eq10.1}
\rho (x,x^{\prime },t)=\stackunder{m}{\sum }w_m\varphi _m(x,t)\overline{\varphi
_m(x^{\prime },t)},
\end{equation}

where $w_m$ is the probability that at the moment of time $t=0$ the
system is in the state $\varphi _m(x,t).$ The function $\varphi _n(x)$ is
the solution of Schr\"{o}dinger equation that satisfies the initial condition

\begin{equation}
\label{eq10.2}
\varphi _m(x)=\left. \varphi _m(x,t)\right| _{t=0},
\end{equation}

At $t=0$ the following density matrix is defined with the help of a set
of wave functions $\varphi _m(x):$

\begin{equation}
\label{eq10.3}
\rho (x,x^{\prime })=\stackunder{m}{\sum }w_m\varphi _m(x)\overline{\varphi
_m(x^{\prime })}.
\end{equation}

Note that the density matrix in the form of (\ref{eq10.3}) was first defined by
Dirac and von Neuman (see \cite{Hart}), the following microcanonical distribution
being introduced for the coefficient $w_m:$

\begin{equation}
\label{eq10.4}
w_m=\exp \left( -\frac{E_m}{kT}\right) ,
\end{equation}

where $E_m$ is the energy of quantum level $m$, $k$ is the Boltzman constant
and $T$ is the temperature of the thermostat.

In particular, it follows from the definition (\ref{eq10.1})-(\ref{eq10.3})
that if at an initial moment of time $t=0$ the system is in the state
$\varphi _m(x,0)$ with probability $w_m$, then the probability for the system
to be in the state $\varphi _m(x,t)$ at the moment of time $t$ will be the same.
One should note that the representation (\ref{eq10.3}) is valid only if the
quantum system weakly interacts with the thermostat.

Below we shall study some relaxation processes that occur with the energy
spectrum of vacuum-immersed quantum harmonic oscillator (QHO), as well as
calculate the thermodynamic potentials of vacuum state when no any limitations
are imposed on the amplitude of interaction between the quantum oscillator
and vacuum.

Because each quantum state $m$ in the problem under consideration is
described by a complex random process, it makes sense to apply here the
thermodynamic description.

{\bf Definition 2.} The partial density matrix is defined to be

\begin{equation}
\label{eq10.5}
\rho _{stc}^{\left( m\right) }=\left( x,t;\left\{ \xi \right\} |x^{\prime
},t^{\prime };\left\{ \xi ^{\prime }\right\} \right) =\left\{ \Psi
_{stc}\left( m|x,t;\left\{ \xi \right\} \right) \overline{\Psi _{stc}\left(
m|x^{\prime },t^{\prime };\left\{ \xi ^{\prime }\right\} \right) }\right\} .
\end{equation}

Let us remember, that the wave functional $\Psi _{stc}\left( m|x,t;\left\{ \xi
\right\} \right) $ in (\ref{eq10.5}) describes QHO state with the frequency

\begin{equation}
\label{eq10.6}
\Omega \left( t\right) =\Omega _{as}+\Omega _{+}\left( t;\left\{
W_{+}\right\} \right) ,\qquad \Omega _{as}=\limfunc{const},
\end{equation}

where the stochastic term of frequency squared have the following form

\begin{equation}
\label{eq10.7}
U_{+}\left( t\right) =2\Omega _{as}\Omega _{+}\left( t;\left\{ W_{+}\right\}
\right) +\Omega _{+}^2\left( t;\left\{ W_{+}\right\} \right)
=F_{0+}+F_{+}\left( t\right) ,\qquad \left\langle U_{+}\left( t\right)
\right\rangle =F_{0+}=\limfunc{const}.
\end{equation}

We shall suppose that 

\begin{equation}
\label{eq10.8}
\left\langle F_{+}\left( t\right) \right\rangle =0,\qquad \left\langle
F_{+}\left( t\right) F_{+}\left( t^{\prime }\right) \right\rangle
=2\varepsilon _{+}\delta \left( t-t^{\prime }\right) .
\end{equation}

{\bf Definition 3.} The mathematical expectation of the stochastic operator
$\hat A\left( x,t;\left\{ \theta \right\} \right) $ in the quantum state $m$
is defined as 

\begin{equation}
\label{eq10.9}
A=\stackunder{t\rightarrow +\infty }{\lim }\left\{ \left. Sp_x\left[
Sp_{\left\{ \xi \right\} }\hat A\rho _{stc}^{\left( m\right) }\right]
\right/ Sp_x\left[ Sp_{\left\{ \xi \right\} }\rho _{stc}^{\left( m\right)
}\right] \right\} .
\end{equation}

where $Sp_{\left\{ x\right\} }$ means the integration along the diagonal line
$x=x^{\prime } $, and $Sp_{\left\{ \theta \right\} }$ denotes respectively the
functional integration along the diagonal $\left\{ \theta \right\} =\left\{ \theta
^{\prime }\right\} $ with measure (\ref{eq7.2})-(\ref{eq7.4}) taking into account
that in the asymptotic subspace $R_{as}$ the following notations were made
$\Omega _0(t)\rightarrow \Omega _{as}$, $F_{0} \rightarrow F_{0+}$, $\gamma
\rightarrow \gamma_{+}=1+\left. F_{0+}\right/ \Omega ^{2}_{as}$, and
$\varepsilon \rightarrow \varepsilon _{+}$ .

{\bf Definition 4.} We shall call the following distribution function
"the nonequilibrium partial" one 

\begin{equation}
\label{eq10.10}
\vartheta \left( \varepsilon _{+},E,t\right) =Sp_x\left\{ Sp_{\left\{ \xi
\right\} }\rho _{stc}^{\left( m\right) }\right\} =Sp_x\left\{ Sp_{\left\{
\theta \right\} }\tilde \rho _{stc}^{\left( m\right) }\left( x,t;\left\{
\theta \right\} |x^{\prime },t^{\prime };\left\{ \theta ^{\prime }\right\}
\right) \right\} ,
\end{equation}

where $E$ is the average value of energy in the quantum state characterized
by the index $m$, $Sp_{\left\{ \theta \right\} }$ standing for the functional
integration along the diagonal line $\left\{ \theta \right\} =\left\{ \theta
^{\prime }\right\} $ but with the measure

\begin{equation}
\label{eq10.11}
D\tilde \mu \left\{ \theta \right\} =\alpha ^{-1}\left( \varepsilon
_{+},\Omega _{as};t\right) \frac{\tilde N_E\left( \varepsilon _{+},\theta
;t\right) }{\tilde N_\Sigma \left( \varepsilon _{+},\theta ;t\right) }D\mu
\left\{ \theta \right\} ,
\end{equation}

where $\overline{N}_E(\varepsilon _{+},\theta ;t)$ is the solution of the
Fokker-Plank equation (\ref{eq6.3})-(\ref{eq6.4}) after replacement of 
$U_0(t)$ by $-E$, $E>0$, and $\overline{N}_\Sigma $, respectively, after
replacement of $U_0(t)$ by $E$. It is easy to see that the ratio
$\left. \overline{N}_E\right/ \overline{N}_\Sigma $ gives the distribution of
quantum states in the vicinity of binding energy in the functional space
$R_{\{\theta _{+}\}}$. It is necessary to stress that the average energy $E$
here is determined with the help of formula (\ref{eq10.9}) and, hence, depends
on the quantum number $m$. After simple integration in (\ref{eq10.10}) with due
regard for (\ref{eq10.11}) and passing to a limit $t\rightarrow +\infty $ we
obtain for the equilibrium distribution function of quantum states

\begin{equation}
\label{eq10.12}
\vartheta ^{\left( m\right) }\left( \varepsilon _{+},E\right)
=\frac{N_E}{N_\Sigma },\qquad E>0,\qquad N_\Sigma =N_{-E},
\end{equation}

where

\begin{equation}
\label{eq10.13}
N_E^{-1}=\pi ^{-1}\varepsilon _{+}^{1/2}\stackrel{\infty }{\stackunder{0}{%
\int }}dzz^{-1/2}\exp \left( -\frac{z^3}{12}+\frac E{\varepsilon
_{+}^{3/2}}z\right) .
\end{equation}

Having in view the asymptotic estimates (\ref{eq6.14})-(\ref{eq6.17}) one can
conclude that when the condition $\left. E\right/ \varepsilon _{+}^{2/3}\ll 1$
is met, the partial distribution function $\vartheta ^{(m)}(\varepsilon _{+},E)$
passes into the microcanonical distribution when the constant $\varepsilon
_{+}=\frac 34E^{\frac 12}kT.$ Owing to that, the integration of stochastic
density matrix (\ref{eq10.5}) over the functional space $R_{\{\theta _{+}\}}$
with the measure (\ref{eq10.11}) and summation over the index $m$ gives the density
matrix in the representation of Dirac and von Neuman 
(\ref{eq10.1})-(\ref{eq10.3}).

Now, having the partial distribution function $\vartheta ^{(m)}(\varepsilon
_{+},E)$ one can determine the thermodynamic potentials of the specific
level $m:$

a) the average internal energy 

\begin{equation}
\label{eq10.14}
{\sf U}^{\left( m\right) }\left( \varepsilon _{+},E\right) =-\partial
_{\varepsilon _{+}}\left\{ \ln \vartheta ^{\left( m\right) }\left(
\varepsilon _{+},E\right) \right\} ,
\end{equation}

b) the free Helmholtzian energy

\begin{equation}
\label{eq10.15}
{\sf F}^{\left( m\right) }\left( \varepsilon _{+},E\right) =-\varepsilon
_{+}^{-1}\ln \vartheta ^{\left( m\right) }\left( \varepsilon _{+},E\right) ,
\end{equation}

c) the entropy

\begin{equation}
\label{eq10.16}
{\sf S}^{\left( m\right) }\left( \varepsilon _{+},E\right) =\varepsilon
_{+}k\left\{ {\sf U}^{\left( m\right) }\left( \varepsilon _{+},E\right) +%
{\sf F}^{\left( m\right) }\left( \varepsilon _{+},E\right) \right\} .
\end{equation}

Having in view the illustration of results we shall carry out the
calculations for vacuum state, i.e., $m=0.$ An appropriate partial
stochastic matrix in the immobile reference will have the form (see Section 4)

\begin{equation}
\label{eq10.17}
\begin{array}{c}
\tilde \rho _{stc}^{\left( 0\right) }\left( x,t;\left\{ \theta \right\} \mid
x^{\prime },t^{\prime };\left\{ \theta ^{\prime }\right\} \right) =\QOVERD(
) {\Omega _{as}}{\pi }^{\frac 12}\exp \left\{ -
\frac{\Omega _{as}}2(x^2+x^{\prime 2})-\right.  \\  \\
\left. -\frac 12\stackrel{t}{\stackunder{-\infty }{\int }}\theta (\tau
)d\tau -\frac 12\stackrel{t^{\prime }}{\stackunder{-\infty }{\int }}\theta
(\tau )d\tau -i\left[ \theta (t)x^2-\theta (t^{\prime })x^{\prime 2}\right]
\right\} ,
\end{array}
\end{equation}

where $\Omega _{as}$ is the permanent frequency of QHO in the asymptotic
space $R_{as}^1.$

Substituting the expression for stochastic potential (\ref{eq2.2}) and stochastic
density matrix (\ref{eq10.17}) to (\ref{eq10.9}) and  making simple calculations
we obtain the following expression for the average value of energy of
"vacuum-oscillator" system in the ground state:

\begin{equation}
\label{eq10.18}
\begin{array}{c}
E^{(0)}\left( \lambda _{+};\Omega _{as}\right) =\frac 12\Omega _{as}J_{+}
\stackrel{\infty }{\stackunder{0}{\int }}dzz^{-3/2}\exp \left( -\frac{z^3}{12%
}-\lambda _{+}z\right) + \\  \\
+\frac 12\Omega _{as}\left\{ 1-
\frac{J_{+}}{\lambda _{+}}\stackrel{\infty }{\stackunder{0}{\int }}%
dzz^{3/2}\exp \left( -\frac{z^3}{12}-\lambda _{+}z\right) \right\} + \\  \\
+i\frac{\Omega _{as}}{2\sqrt{\lambda _{+}}}J_{+}\stackrel{\infty }{%
\stackunder{0}{\int }}dzz^{1/2}\exp \left( -\frac{z^3}{12}-\lambda
_{+}z\right) ,
\end{array}
\end{equation}

where the notation was made

$$
J_{+}^{-1}=\pi ^{1/2}\varepsilon _{+}^{-1/3}\stackrel{\infty }{\stackunder{0}{%
\int }}dzz^{-1/2}\exp \left( -\frac{z^3}{12}-\lambda _{+}z\right) .
$$

Note that the expression (\ref{eq10.18}) was obtained taking into account that
$\left\langle F\right\rangle _{\{W_{+}\}}=0.$ One can see that the first term
in (\ref{eq10.18}) is diverging that corresponds to the infinite energy of
physical vacuum. The second term corresponds to the energy of oscillator in the
ground state that is shifted as a result of interaction with the nonrelativistic
vacuum (see FIG.3)

\begin{equation}
\label{eq10.19}
\begin{array}{c}
E_{osc}^{(0)}\left( \lambda _{+};\Omega _{as}\right) =\frac 12\Omega _{as}\left\{
1-\frac 1{\lambda _{+}}\left. \left[ \left( \partial _\alpha \ln A\left(
-\lambda _{+}+\alpha \right) \right) ^2+\partial _\alpha ^2\ln A\left(
-\lambda _{+}+\alpha \right) \right] \right| _{\alpha =0}\right\} , \\
\\
A\left( -\lambda _{+}+\alpha \right) =Ai^2\left( -\lambda _{+}+\alpha
\right) +Bi^2\left( -\lambda _{+}+\alpha \right) ,
\end{array}
\end{equation}

It is noteworthy that the second term in (\ref{eq10.19}) is an analogue of the
Lamb shift of the energy level, that is well  known from the QED \cite{Thirring}.
The third term in (\ref{eq10.18}) corresponds to a broadening of the ground
level and, hence, is inverse proportional to its decay time

\begin{equation}
\label{eq10.20}
\Delta t^{(0)}\sim 2\frac{\sqrt{\lambda _{+}}}{\Omega _{as}}\left. \left\{
\partial _\alpha \ln A\left( -\lambda +\alpha \right) \right\} ^{-1}\right|
_{\alpha =0}.
\end{equation}

Now, based on the formulae (\ref{eq10.12})-(\ref{eq10.16}), one can write an
expression for the entropy of QHO in the $m$-th quantum state immersed in the
physical vacuum 

\begin{equation}
\label{eq10.21}
\begin{array}{c}
S^{\left( m\right) }\left( \varepsilon _{+},E_{osc}\right) =-k\varepsilon _{+}%
\left\{ \partial _\alpha \ln 
  \left[ A%
    \left( -\frac{E_{osc}}{\varepsilon _{+}^{2/3}}+\alpha
    \right) A%
    \left( \frac{E_{osc}}{\varepsilon _{+}^{2/3}}+\alpha
    \right)
  \right] +%
\right.
\\ \\
\left.
  \left. +\varepsilon _{+}^{-1}\ln
    \left[
      \left. A%
        \left( \frac{E_{osc}}{\varepsilon _{+}^{2/3}}%
        \right) 
      \right/ A%
      \left( -\frac{E_{osc}}{\varepsilon _{+}^{2/3}}%
      \right) 
    \right] 
  \right\} 
\right| _{\alpha =0}.
\end{array}
\end{equation}

To obtain an expression for the entropy of oscillator in the ground state
one has to make a substitution $E_{osc}\rightarrow E_{osc}^{(0)}$ in the
formula (\ref{eq10.21})

\begin{equation}
\label{eq10.22}
\begin{array}{c}
S^{\left( 0\right) }\left( \beta _{+}^{\left( 0\right) }\right) =k\left\{
\beta _{+}^{\left( 0\right) }\left[ \frac{%
A_{1/2}\left( -\beta _{+}^{\left( 0\right) }\right) }{A_{-1/2}\left(
-\beta _{+}^{\left( 0\right) }\right) }+\frac{A_{1/2}\left( \beta
_{+}^{\left( 0\right) }\right) }{A_{-1/2}\left( \beta _{+}^{\left( 0\right)
}\right) }\right] +\ln \frac{A_{-1/2}\left( \beta _{+}^{\left( 0\right)
}\right) }{A_{-1/2}\left( -\beta _{+}^{\left( 0\right) }\right) }\right\} ,
\end{array}
\end{equation}

where the notations were made

\begin{equation}
\label{eq10.23}
A_p\left( q\beta _{+}^{\left( 0\right) }\right) =\stackrel{\infty }{%
\stackunder{0}{\int }}dzz^p\exp \left( -\frac{z^3}{12}+q\beta _{+}^{\left(
0\right) }z\right) ,\qquad \beta _{+}^{\left( 0\right) }=\frac{E^{\left(
0\right) }}{\varepsilon _{+}^{2/3}}.
\end{equation}


\section{Conclusions}

At present the three following major schemes for the quantum chaos initiation
are discussed:

a) the dynamical one, when the classical analogue of the quantum object
under study is a nonintegrable system \cite{Gutzwiller}. An evident example of
such a case is the three-body problem \cite{Minsk}-\cite{Bogdanov};

b) the measurement problem comprising the issues of mesophysics, of the
irreversibility problem and quantum jumps \cite{Von Oppen},
\cite{Balian}-\cite{Weizsacker};

c) randomness, and, therefore, the irreversibility as the basic properties
of the physical world, owing to which the really existing object,-the
physical vacuum (see \cite{Birrell}), is included in theoretical consideration.

The objective of the present paper was to investigate the third case. The main
point here is the idea that the quantum object and the physical vacuum are
considered to make a joint system described in the framework of single equation.
Since such a system has infinite number of the degrees of freedom, it is
convenient to mathematically formulate the problem in the framework of complex
SDE in the extended space $\Xi =R^1\otimes R_{\left\{ \xi \right\} }$. In this
case the system "quantum object+thermostat" will be described by the wave
functional, - a complex probabilistic process, and not by the wave function.
As a specific example, the problem of one-dimensional QRHO motion in the Euclidean
space $R^1$ has been considered in the work. It was shown that in the complex
SDE (\ref{eq2.1})-(\ref{eq2.2}) the variables were separated by means of
Langevin-type standard real nonlinear SDE and the solutions for the wave
state were obtained in the form of complex orthonormalized functionals $\Psi
_{stc}^{(+)}\left( n\mid x,t;\left\{ \xi \right\} \right) $ in the space $%
L_2\left( R^1\otimes R_{\left\{ \xi \right\} }\right) $. For nonlinear
Langevin SDE the corresponding Fokker-Plank equation has been obtained and
with its help a complete positive measure for the functional space was
constructed. This last fact permits the construction of uniformely converging
representation for the averaged wave function $\Psi _{br}^{(+)}\left( n\mid
x,t\right) $ of moving QRHO on the conventional space-time continuum $(R^1,t)
$ by means of functional integration of the wave process $\Psi
_{stc}^{(+)}\left( n\mid x,t;\left\{ \xi \right\} \right) $ in the space $%
R_{\left\{ \xi \right\} }$. The obtained mixed continual-undulatory
representation unites two concepts that at the first sight seem 
contradictory: the quantum analogue of the Arnold transformations
\cite{Schuster} that excludes the rise of chaos inside the given trajectory
beam, i.e., forbids any changes of beam topology, and the functional integration
that allows for the contribution of various topological pipes and thereby
assists to the generation of the chaos. With the help of wave functionals
$\Psi _{stc}^{(+)}\left( n\mid x,t;\left\{ \xi \right\} \right) $ an expression
for the elements of transition matrices (see (\ref{eq8.3})) was constructed.
It was shown that the stochastic matrix formed in such a way is unitary $%
S^{stc}(S^{stc})^{+}=\mathop{\rm I}$. The generalized transition matrix $%
S^{br}$ for QRHO was obtained by averaging of the stochastic matrix $S^{stc}$
in the $R_{\left\{ \xi \right\} }$ space and it was shown to be generally
nonunitary. In other words, in the theoretical scheme in question the
quantum oscillator is an open system in the space-time continuum $(R^1,t)$
and here the conservation laws are violated. Nevertheless, when the
oscillator weakly interacts with the thermostat the transition matrix is
simplified and turns to be unitary, $S_W^{br}(S_W^{br})^{+}=\mathop{\rm I}$,
i.e., the conservation laws are effective again.

In the present paper a detailed analysis of the probability of
"vacuum-vacuum" transition $\Delta _{0\rightarrow 0}^{br}\left(
\lambda ,\rho \right) $ is given in terms of $\lambda $ that characterizes
random fluctuations of the thermostat before and after the crossover, as
well as in terms of barrier reflection coefficient $\rho $ in the relevant
problem of the one-dimensional quantum mechanics. It was shown that for
relatively strong interactions of the oscillator with the thermostat, i.e.,
for relatively small parameter $\lambda $, the $\rho$-dependence of the
probability is nonmonotonic (see FIG.2).

Based on the example of vacuum-immersed QHO, a novel formalism of quantum
statistical mechanics has been developed in the framework of density matrix
approach (\ref{eq10.5}). It was shown that in the limit of weak
oscillator-vacuum interaction, i.e., at $\lambda _{+} \rightarrow \infty $
it can be reduced to the Dirac and von Neuman density matrix. In the framework
of novel approach the relaxation effects related to the energy spectrum of the
oscillator were studied in detail. In particular, analytical expressions for
the energy of ground state with broadening and shift (analogous to the well
known Lamb shift from the Quantum Electrodynamics) as a function of constant
$\lambda _{+}$, and for the entropy of the ground (vacuum) state of the
oscillator were obtained.

We should like to draw attention to one very important property of the
quantum representation under consideration, viz., that it permits to relate
the domain of quantum chaos with that of classical chaos \cite{Bogdanov}, the
transition to the classical chaotic dynamics taking place in the limit
$\hbar \rightarrow 0$ when $\varepsilon \neq 0$. In case when $\varepsilon
\rightarrow 0$ and $\hbar \neq 0$ SDE (\ref{eq2.1}) goes into the usual
Schr\"{o}dinger equation and describes a reversible quantum process. The
classical theory of reversible processes is obtained when one tends to zero
$\hbar \rightarrow 0$ and $\varepsilon \rightarrow 0$.

An explicit relation between the quantum mechanics and quantum statistical
thermodynamics was also established to specify, in particular, the range of
applicability of the statistical matrix of Dirac and von Neuman.

In conclusion we should like to note that one can successfully develop the
quantum mechanics in the framework of complex SDE (\ref{eq2.1}) both for the
model of multidimensional oscillator and also for other exactly solved
nonstationary problems of quantum mechanics. Its covariant generalization is
also not difficult, and we postpone the detailed discussions of these
problems to our later publications.

\begin{figure}
\setlength{\unitlength}{1 cm}
\begin{picture}(18,18)(0,0)
\put(0,18){\special {em:graph 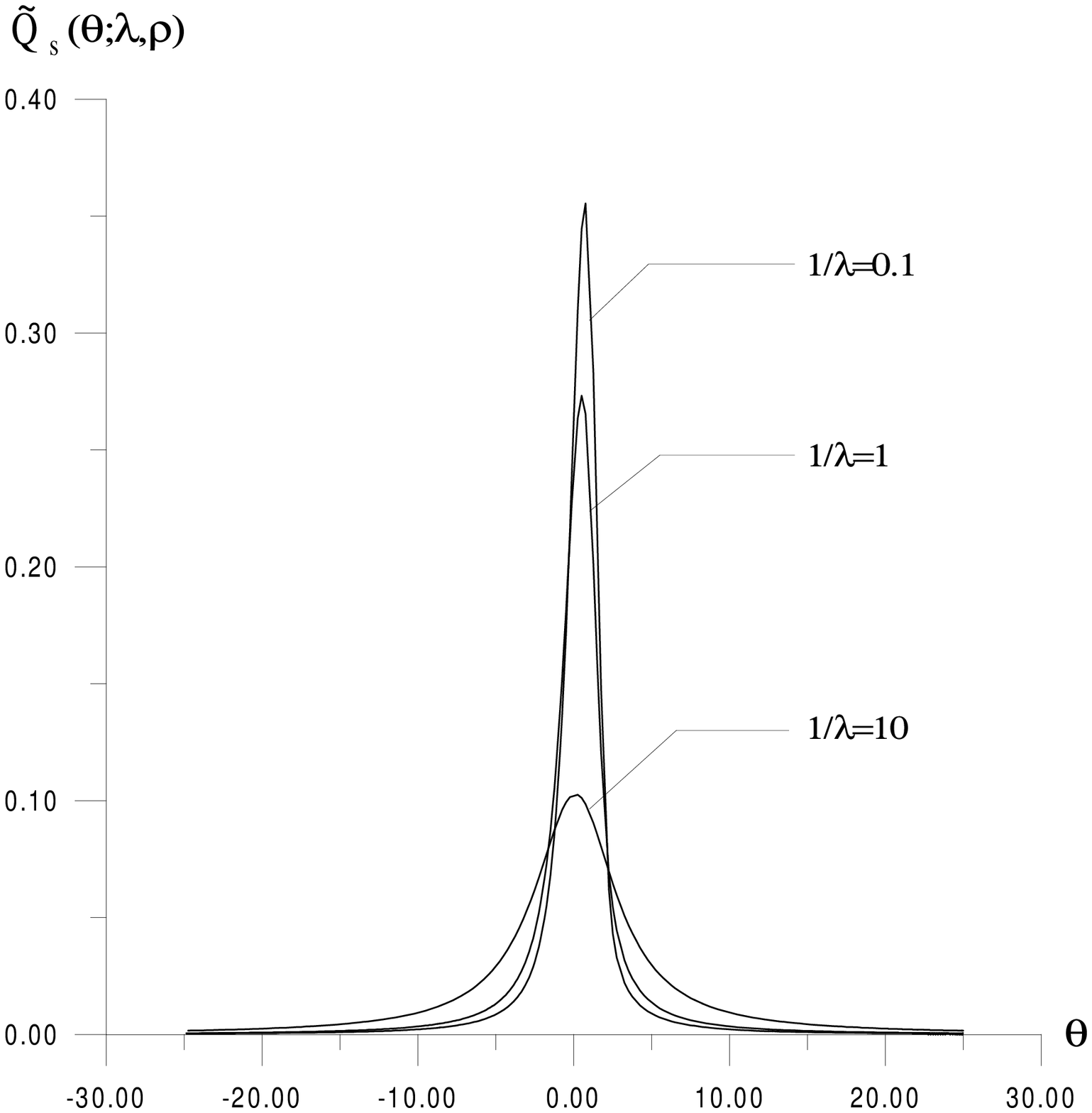}}
\end{picture}
\caption{The dependence of the distribution  of stationary process
$\tilde Q_s\left( \lambda ;\bar \theta \right) $ of over $\bar \theta $ 
for different values of parameter $1/\lambda \sim \varepsilon $.}
\end{figure}

\begin{figure}
\setlength{\unitlength}{1 cm}
\begin{picture}(18,18)(0,0)
\put(0,18){\special {em:graph 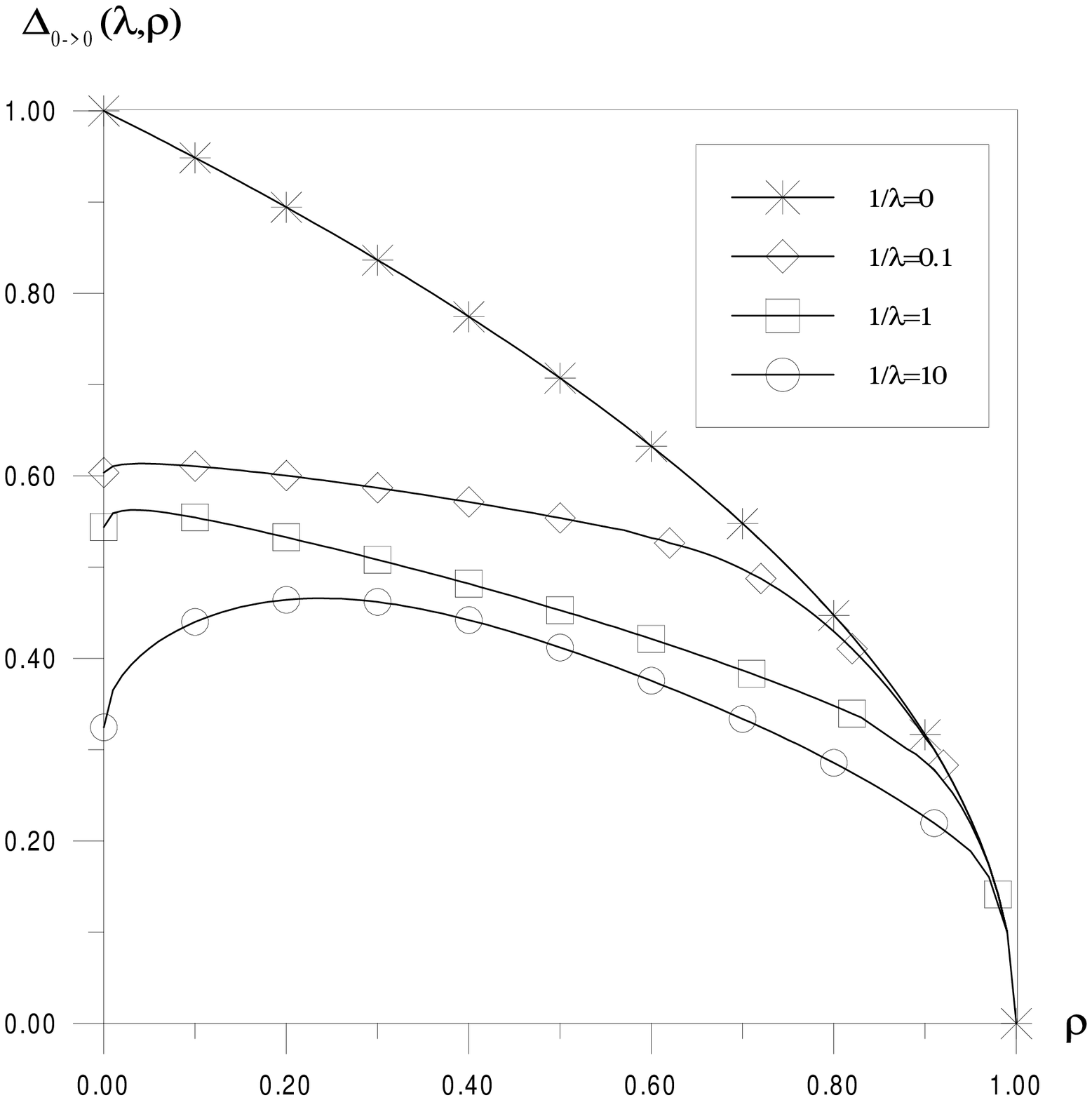}}
\end{picture}
\caption{"Vacuum-vacuum" transition probability in dependence of
$\lambda $ and $\rho $.}
\end{figure}

\begin{figure}
\setlength{\unitlength}{1 cm}
\begin{picture}(18,16)(0,0)
\put(0,16){\special {em:graph 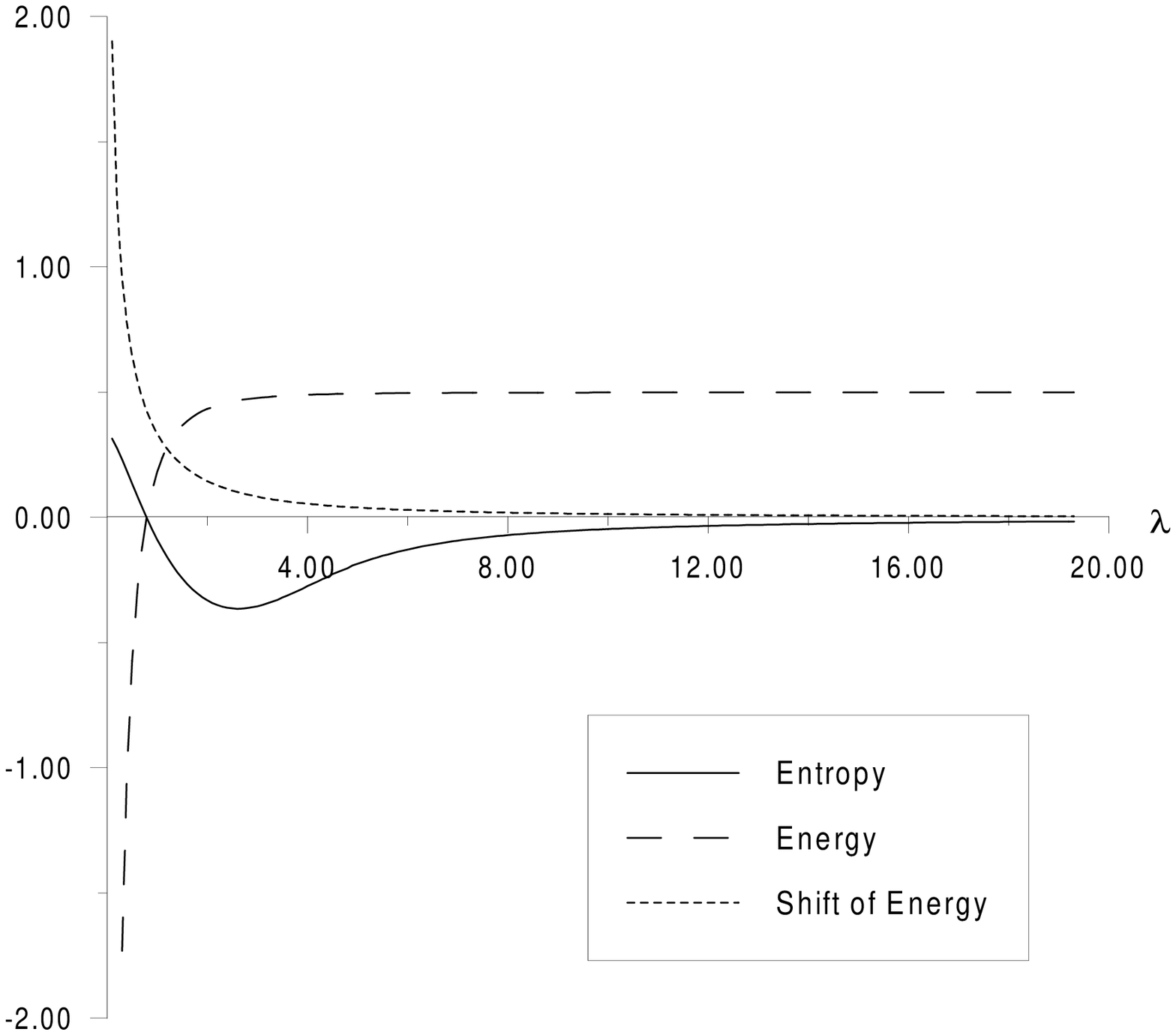}}
\end{picture}
\caption{Dependence of oscillator ground state energy, its shift and entropy
of vacuum over the parameter $\lambda _{+}$.}
\end{figure}

\end{document}